\documentclass[12pt,twocolumn]{article}

\usepackage{arxiv}
\usepackage{lmodern}

\usepackage[utf8]{inputenc} 
\usepackage[T1]{fontenc}    
\usepackage{hyperref}       
\usepackage{url}            
\usepackage{booktabs}       
\usepackage{amsfonts}       
\usepackage{nicefrac}       
\usepackage{microtype}      
\usepackage{lipsum}
\usepackage{comment}
\usepackage{subcaption}
\usepackage{graphicx}
\usepackage{textcomp}
\usepackage{soul}           
\usepackage{amsmath}
\usepackage{bm}
\usepackage[sorting=none,giveninits,style=numeric-comp,doi=false,isbn=false,url=false,date=year,eprint=false,maxnames=5]{biblatex}
\renewbibmacro{in:}{}
\AtEveryBibitem{\clearfield{pages}}
\AtEveryBibitem{\clearfield{number}}

\usepackage{chngcntr}
\usepackage[font=small,labelfont=bf,textfont=it]{caption}
\usepackage{enumitem}
\usepackage[table,xcdraw]{xcolor}

\def\mycmd{0} 

\title{Deep learning the slow modes for rare events sampling
}
\author{
  Luigi Bonati\thanks{luigi.bonati@iit.it}
  \vspace{0.5em}\\
  Department of Physics, ETH Zurich, 8092 Zurich, Switzerland\\
  Atomistic Simulations, Italian Institute of Technology, 16163 Genova, Italy\\
   \And GiovanniMaria Piccini
  \vspace{0.5em}\\
  Basic \& Applied Molecular Foundations, Physical and Computational Sciences Directorate, \\Pacific Northwest National Laboratory, Richland, WA 99352, USA\\
  Istituto Eulero, Università della Svizzera italiana, 6900 Lugano, Switzerland\\
   \AND
  Michele Parrinello\protect\thanks{michele.parrinello@iit.it}
  \vspace{0.5em}\\
  Atomistic Simulations, Italian Institute of Technology, 16163 Genova, Italy\\
} 

\defineheader{\small Bonati \textit{et al.}, Deep learning the slow modes for rare events sampling}

\begin{document}
\twocolumn[
  \begin{@twocolumnfalse}
    \maketitle
        \begin{abstract}
The development of enhanced sampling methods has greatly extended the scope of atomistic simulations, allowing long time phenomena to be studied with accessible computational resources.  Many such methods rely on the identification of an appropriate set of collective variables. These are meant to describe the system’s modes that most slowly approach equilibrium. Once identified, the equilibration of these modes is accelerated by the enhanced sampling method of choice. An attractive way of determining the collective variables is to relate them to the eigenfunctions and eigenvalues of the transfer operator. Unfortunately, this requires knowing the long-term dynamics of the system beforehand, which is generally not available. However, we have recently shown that it is indeed possible to determine efficient collective variables starting from biased simulations. In this paper, we bring the power of machine learning and the efficiency of the recently developed on-the-fly probability enhanced sampling method to bear on this approach. The result is a powerful and robust algorithm that, given an initial enhanced sampling simulation performed with trial collective variables or generalized ensembles, extracts transfer operator eigenfunctions using a neural network ansatz and then accelerates them to promote sampling of rare events. To illustrate the generality of this approach we apply it to several systems, ranging from the conformational transition of a small molecule to the folding of a mini-protein and the study of materials crystallization.

\keywords{Enhanced sampling $|$ Collective variables $|$ Machine learning} 
\vspace{1em}
\end{abstract}
  \end{@twocolumnfalse}
  ]

\if\mycmd0
  Atomistic simulations
  \else
  \dropcap{A}tomistic simulations
 \fi
and in particular molecular dynamics (MD), play an important role in several fields of science, serving as a virtual microscope that is of great help in the study of physical, chemical and biological processes.  
However, any time the free energy barrier between metastable states is large relative to the thermal energy $k_B T$ transitions between states become rare events, taking place on time scales too long to be simulated by standard methods~\cite{Peters2017}.
This severely hampers the study of many important phenomena, such as phase transitions, chemical reactions, protein folding and ligand binding. 

To alleviate this problem, different advanced sampling techniques have been developed. A large family of these methods relies on the identification of a small set of collective variables (CVs) $\bm{s}=\bm{s}(\bm{R})$, that are functions of the system atomic coordinates $\bm{R}$. 
In all these approaches an external bias potential $V\left(\bm{s}(\bm{R})\right)$  is added to the system in order to enhance the $\bm{s}(\bm{R})$ fluctuations~\cite{Valsson2016}. If the CVs are able to activate the slowest degrees of freedom involved in the state-to-state transitions this procedure results in an enhanced  sampling of the transition state. This in turn leads to an increase in the frequency with which rare events are sampled. 
Different ways of constructing appropriate bias potentials have been suggested. Examples are umbrella sampling~\cite{Torrie1977,Mezei1987AdaptiveBias}, hyperdynamics~\cite{Voter1997AcceleratedEvents}, metadynamics~\cite{Laio2002}, variational enhanced sampling~\cite{Valsson2014,Bonati2019NeuralSampling}, Gaussian mixture-based enhanced sampling~\cite{Debnath2020GaussianDynamics} and on-the-fly probability enhanced sampling (OPES) \cite{Invernizzi2019b}.

Regardless of the method used, identifying appropriate collective variables is a  crucial requisite for a  successful enhanced sampling simulation~\cite{Barducci2011Metadynamics, Bussi2020UsingLandscapes}.
Ideally one would choose the CVs solely on a physical and chemical basis. However, especially for complex systems, this can be rather cumbersome.
For this reason, a number of data-driven approaches and signal analysis methods have been proposed for CV construction~\cite{Sidky2020MachineSimulation,Wang2020MachineSimulations,Noe2017CollectiveMethods}.
Some of these methods can be applied when the metastable states involved in the rare event are known beforehand~\cite{Mendels2018, Sultan2018, Brandt2018MachineCoordinates}, such as the folded and the unfolded states of a peptide or the reactants and products of a reaction. A line of attack in  these cases has been to collect a number of configurations from short unbiased MD runs in the different metastable states and use these data to train a supervised classification algorithm. The classifier is then used as a  CV. Our group has also contributed to this literature and developed an approach named harmonic linear discriminant analysis~\cite{Mendels2018, Piccini2018} that derives from Fisher's linear discriminant analysis (LDA). Later we have further improved this method by applying  a non-linear version of LDA (Deep-LDA)~\cite{Bonati2020Data-DrivenSampling}. The greater flexibility provided by the neural network architectures is of great help in dealing with complex problems~\cite{Rizzi2021TheInteraction,Karmakar2021CollectiveCrystallisation}. These methods have proven to be successful in spite of the fact that they do not necessitate prior knowledge of reaction paths or transition states. 

Clearly, if we had access to the transition dynamics we could further improve the CV effectiveness  by making use of this dynamical information. 
To this purpose, several methods have been suggested to extract CVs from reactive simulations in which the system translocates spontaneously from one metastable state to another. Among all these methods those based on the variational approach to conformational dynamics (VAC)~\cite{Naritomi2011SlowMotions,Prinz2011MarkovValidation,Perez-Hernandez2013,Schwantes2013ImprovementsNTL9,Wu2017VariationalSimulations,M.Sultan2017a,McCarty2017c,Hernandez2018,Wehmeyer2018b,Mardt2018,Chen2019}, both in its linear and non linear versions, are of particular relevance here. 
In Ref.~\cite{McGibbon2017IdentificationDynamics} it has been argued that the resulting variables are natural reaction coordinates since they a) perform a dimensionality reduction, b) are determined by the sampling dynamics and c) are maximally predictive of the system evolution.
Furthermore, an interesting feature of these CVs is that they measure the progress along any pathway connecting the metastable states, rather than focusing on a single path~\cite{McGibbon2017IdentificationDynamics}. Hence these variables can be of great help both to understand and to enhance MD simulations. 
However, a difficulty in using VAC-generated CVs is that it becomes superfluous to perform enhanced sampling if unbiased reactive trajectories are already available. Thus one is in a chicken-and-egg situation: to find good CVs one needs to collect unbiased state-to-state transitions, but to promote transitions good CVs are needed~\cite{Rohrdanz2013DiscoveringReactions}. 

A solution to this conundrum may come from an iterative approach, in which the CVs are computed using data generated in a previous enhanced sampling simulation~\cite{Tiwary2016,Chen2018,Demuynck2018ProtocolFrameworks,Ribeiro2018,Zhang2018UnfoldingSampling,Wang2019,Rydzewski2021MultiscaleSampling,Tsai2021SGOOP-d:Simulations,Belkacemi2021ChasingTrajectories}, even if this initial run is far from optimal.
In our group we have followed this strategy and modified the VAC protocol to identify the slow modes from biased trajectories~\cite{McCarty2017c,Yang2018}. In this way we can identify the modes that hinder convergence and enhance their sampling. 

Here we generalize the approach of Ref.~\cite{McCarty2017c} in two ways. First, we employ a non-linear variant of VAC, which greatly increases its variational flexibility. Second, we propose new strategies for the collection of the initial trajectories, such as sampling generalized ensembles rather than using trial CVs, and for making full use of the information gathered during the initial trajectory.
We also employ OPES to construct the bias, which has several advantages over metadynamics and other methods. These improvements lead to a general procedure that is proven to be effective in the study of a variety  of rare events.

We organize the structure of this paper as follows. First, we give a brief account of the VAC theory, highlighting the points that are most relevant to our work. Then, we discuss how we can use neural networks as trial functions for the variational principle and how to adapt it when starting from enhanced sampling simulations. We initially test our method on the didactically informative example of the alanine dipeptide and then move on to more substantial applications such as folding a small protein and studying a crystallization process.

\subsection*{Collective variables as eigenfunctions of the transfer operator}

A molecular dynamics simulation can be seen as a dynamical process that takes a density distribution $p_t(\bm{R})$ at time $t$ and evolves it towards the equilibrium Boltzmann one $\mu \left( \bm{R}\right)= \frac{e^{-\beta U(\bm{R}) }}{\int d \bm{R}\ e^{-\beta U(\bm{R}) } }$.
Here $\beta$ is the inverse temperature and $U(\bm{R})$ the interaction potential.
An analysis of sampling dynamics can be done by studying the properties of the transfer operator $\mathcal{T}_\tau$. We assume the dynamics to be reversible, thus satisfying the detailed balance condition. In the following, we quote some of the properties of $\mathcal{T}_\tau$ and refer the interested reader to the literature for a more formal discussion~\cite{Prinz2011MarkovValidation,Noe2013ASystems}.

The transfer operator is defined by its action on the deviation of the probability distribution $p_{t}\left( \bm{R}\right)$ from its Boltzmann value $\mu(\bm{R})$ as measured by $u_{t}\left( \bm{R}\right) =\frac{p_{t}\left( \bm{R}\right) }{\mu \left( \bm{R}\right) }$:

\begin{align}
    &u_{t+\tau }\left( \bm{R}\right) =\mathcal{T}_\tau \circ u_{t}\left( \bm{R}\right) \\
    &=\dfrac{1}{\mu \left( \bm{R}\right) }\int d\bm{R}'\ \mathcal{P}\left( \bm{R}_{t+\tau }=\bm{R} | \bm{R}_{t}=\bm{R}'\right) u_{t}\left( \bm{R}'\right) \mu \left( \bm{R}'\right)
\end{align}

The action of the transfer operator depends both on the equilibrium distribution $\mu$ and the transition probability $\mathcal{P}$ which is a property of the sampling dynamics. Thus, we will have different operators even when sampling the same equilibrium distribution, e.g. with standard or enhanced sampling molecular dynamics.

The transfer operator $\mathcal{T}_\tau$ is self-adjoint with respect to the Boltzmann measure. This implies that its eigenvalues $\{\lambda_i\}$ are real and that its eigenfunctions $\{\Psi_i(\bm{R})\}$ form an orthonormal basis:
\begin{equation}
     \mathcal{T}_\tau \circ \Psi_i \left( \bm{R}\right) = \lambda_i \Psi_i \left( \bm{R}\right)
\end{equation}
where the orthonormality condition reads:
\begin{equation}
    \langle \Psi_i\ , \Psi_j \rangle _{\mu } = \int d\bm{R}\ \Psi_i\left( \bm{R}\right) \Psi_j\left( \bm{R}\right) \mu \left( \bm{R}\right) = \delta_{ij}
    \label{eq:ortho}
\end{equation}
Furthermore, its eigenvalues are positive and bounded from above : $\lambda_0  = 1 > \lambda_1\ge...\ge\lambda_i\ge.. $. In particular, the eigenfunction corresponding to the highest eigenvalue $\lambda_0 = 1$ is the function $\Psi_0=1$. This trivial solution correspond to the fact that the Boltzmann distribution is the fixed point of $\mathcal{T}_\tau$.
In fact, if we apply $k$ times $\mathcal{T}_\tau$ to a generic density $\psi_t$ we find 
\begin{equation}
    \begin{aligned}
    \psi_{t+k\tau }\left( \bm{R}\right) &=\mathcal{T}_\tau ^{k} \circ \psi_{t}\left( \bm{R}\right)\\
    &=\sum _{i}\langle \Psi _{i}\ , \psi_{t} \rangle_\mu\ \lambda_i^k\Psi _{i}\left( \bm{R}\right) 
    \end{aligned}
\end{equation}
from which we see that if we let $k \to \infty$ only the contribution coming from the $\lambda_0=1$ eigenfunction survives.
The eigenvalues can be reparametrized  as $\lambda_i= e^{-\tau/t_i}$ where $t_i$ is an implied timescale measuring the decay time of the i-th eigenfunction.
Thus, the leading eigenvalues are to be associated with the longest implied timescales, meaning that their corresponding eigenfunctions have a slow relaxation towards the equilibrium. For this reason, the first eigenfunctions are good CV candidates since they describe the slow dynamical process that we need to accelerate.

Of course in a multidimensional system there is no chance of exactly diagonalizing the transfer operator. Nonetheless, we can use a variational approach akin to the Rayleigh–Ritz principle in quantum mechanics. In the present context the variational principle reads 
\begin{equation}
    \lambda_i \ge 
    \frac
    {\langle \tilde{\psi}_i\ , \mathcal{T}_\tau\circ\tilde{\psi}_i \rangle_\mu}
    {\langle \tilde{\psi}_i\ , \tilde{\psi}_i \rangle_\mu} 
    = \dfrac{\langle \tilde{\psi }_{i}\left( \bm{R}_{t}\right) \tilde{\psi }_{i}\left( \bm{R}_{t+\tau }\right) \rangle  }{\langle \tilde{\psi }_{i}\left( \bm{R}_{t}\right) \tilde{\psi }_{i}\left( \bm{R}_{t}\right) \rangle } = \tilde{\lambda}_i
\end{equation}
where $\tilde{\lambda_i}$ are lower bounds for the true eigenvalues and $\tilde{\psi}_i$ are variational eigenfuctions satisfying the orthogonality condition: $\langle \tilde{\psi}_i\ , \tilde{\psi}_j \rangle_\mu = \langle \tilde{\psi }_{i}\left( \bm{R}_{t}\right) \tilde{\psi }_{i}\left( \bm{R}_{t}\right) \rangle = \delta_{ij}\quad\forall j=0,...,i-1$.
The equality holds only when  $\tilde{\psi}_i$ coincides with the exact eigenfunctions.  
In addition, we have used the  property that the matrix elements of $\mathcal{T}_\tau$ can be written in terms of time-correlation function~\cite{Perez-Hernandez2013}. Thus they can be straightforwardly computed from the sampling trajectories.

\subsection*{Time lagged independent component analysis}
In order to solve the variational problem we have to choose a set of trial functions. 
As a first step we select a set of $N_d$ descriptors $\{d_j(\bm{R})\}$ and build the variational eigenfunctions $\tilde{\psi_i}(\bm{R})$  as a linear combination of them: 
\begin{equation}
    \tilde{\psi }_{i}\left( \bm{R}\right) =\sum _{j=1} ^{N_d} \alpha _{ij}d_{j}\left( \bm{R}\right)
\end{equation}
where the expansion coefficients are the variational parameters. 
This amounts to applying to the problem the time-lagged indipendent component analysis (TICA) method \cite{Naritomi2011SlowMotions,Perez-Hernandez2013,Schwantes2013ImprovementsNTL9}, a signal analysis technique that, given a set of variables, aims at finding the linear combination for which their autocorrelation is maximal. 
By imposing the variational functions to have zero mean we ensure that they are orthogonal to the trivial $\psi_0(\bm{R})=1$ solution.
As in quantum mechanics, finding the variational solution to a problem in which a trial wave function is expressed as a linear expansion leads to solving an eigenvalue problem~\cite{Noe2013ASystems}. Recalling that the matrix elements of the transfer operator can be expressed as time correlation functions, the generalized eigenvalue problem can be written as  
\begin{equation}
    C\left( \tau \right) \alpha_{i}=\tilde{\lambda _{i}}C\left( 0\right) \alpha_i
    \label{eq:tica}
\end{equation}
where 
\begin{equation}
    \begin{aligned}
    C_{ij}\left( \tau \right) &=\langle d_{i}\left( \bm{R}_{t}\right)
     d_{j}\left( \bm{R}_{t+\tau }\right) \rangle \\
    C_{ij}\left( 0\right) &=\langle d_{i}\left( \bm{R}_{t}\right) d_{j}\left( \bm{R}_{t}\right) \rangle.\\
    \end{aligned}
\end{equation}

\subsection*{A neural network ansatz for the basis functions}

Instead of using a predetermined set of descriptors as basis functions, as done in TICA, we employ a neural network (NN) to learn the basis functions through a non-linear transformation of the descriptors in a lower dimensional space.
In this way, we can exploit the flexibility of neural networks to drastically improve the variational power of ansatz functions, and at the same time extend the VAC method to the case of a large number of descriptors. 

\begin{figure}[t]
\includegraphics[width=\linewidth]{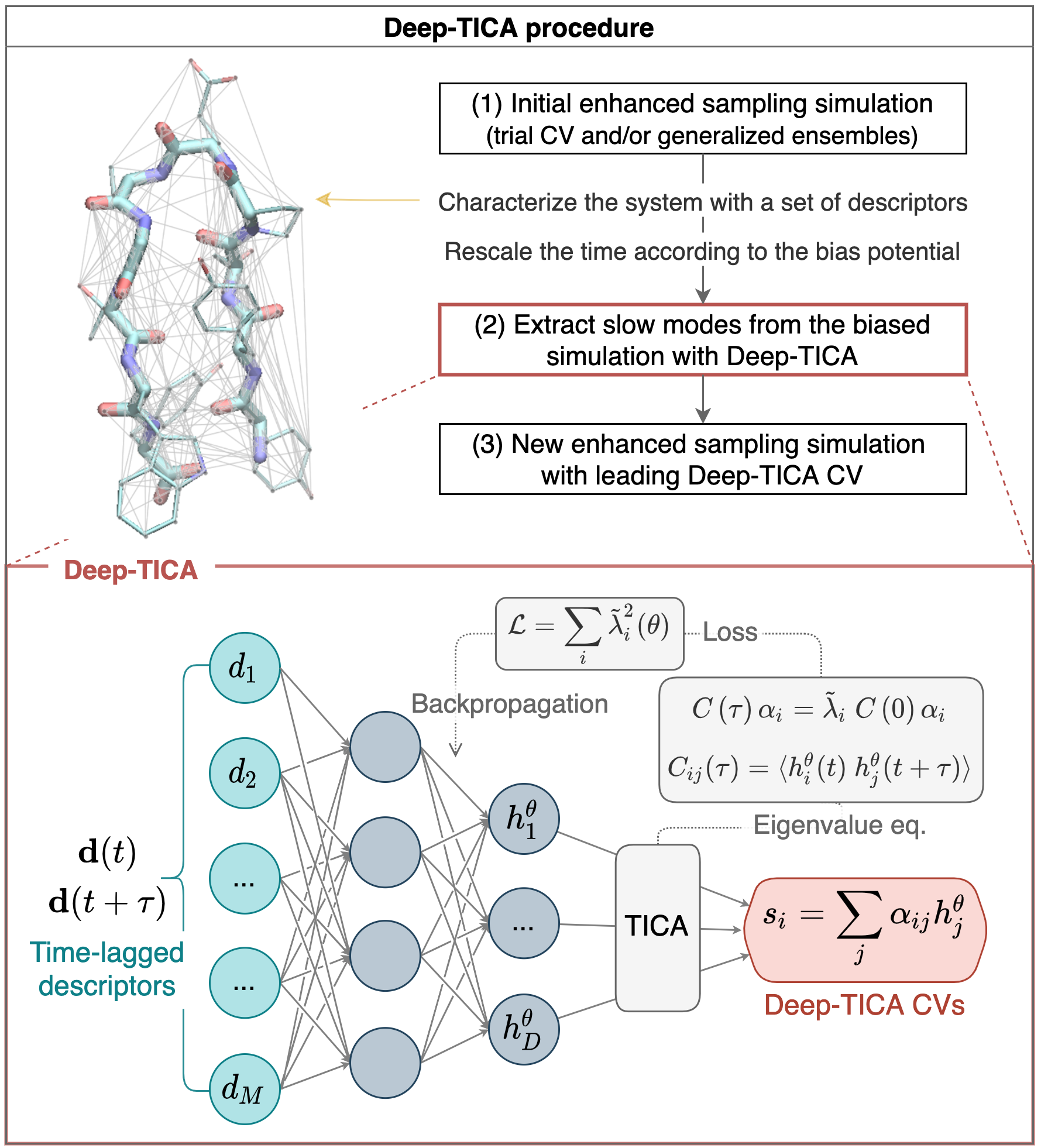}
\centering
\caption{(top) Deep-TICA protocol used in this paper. On the left the mini-protein chignolin is shown, with lines denoting pairwise distances used as descriptors. (bottom) Neural network architecture and optimization details of Deep-TICA CVs. }
\label{fig:scheme}
\end{figure}

The  architecture of the NN follows here the implementation of Ref.~\cite{Chen2019}, where descriptors $\bm{d}(t)=\bm{d}(\bm{R}_t)$ and $\bm{d}(t+\tau)=\bm{d}(\bm{R}_{t+\tau})$ are fed one after the other into a fully-connected neural network parametrized by a set of parameters $\theta$, obtaining as outputs the corresponding latent variables $\bm{h}_\theta\left(\bm{d}(t)\right)$ and $\bm{h}_\theta\left(\bm{d}(t+\tau)\right)$, respectively (see Fig.~\ref{fig:scheme}).
The average of the latent variables $\bm{h}_\theta$ is subtracted to obtain mean free descriptors. Then, these values are used to compute the time-lagged covariance matrices, from which the eigenvalues $\tilde{\lambda}_i (\theta)$ and the corresponding eigenfunctions are obtained as solution of Eq.~\ref{eq:tica}.
This information is used to optimize the parameters of the NN as to maximise the first D eigenvalues, by minimizing the following loss function with gradient descent methods:
\begin{equation}
    \mathcal{L} = - \sum_{i=1} ^{D} \tilde{\lambda}_i ^2 (\theta)
\end{equation}
which corresponds also to the so-called VAMP-2 score \cite{Mardt2018}. 

This architecture is an end-to-end framework that takes as input a set of descriptors and returns as output the few TICA eigenfunctions of interest. 
Since these CVs are given by the combination of NN basis functions and the TICA method, we will refer to them in the following as Deep-TICA CVs. 

\subsection*{Extending TICA to enhanced sampling simulations}

Most of the developments and applications of VAC have been focused on the analysis of long unbiased MD runs. Our purpose is different since we want to identify  the slowest dynamical processes from enhanced sampling simulations. 
The application of an external potential can be seen as an importance sampling technique that samples a modified probability  distribution in which the transition rate is accelerated. To recover the equilibrium properties over the Boltzmann distribution one needs to perform the reweighting procedure~\cite{Valsson2016,Wu2017VariationalSimulations}. 
When the bias is in a quasi-static regime, the expectation value of any operator $\langle O \rangle$ can be written as:
\begin{equation}
    \langle{O \left(\bm{R}\right) }\rangle = \frac{\langle{O\left(\bm{R}\right)\ e^{\beta V\left(\bm{s}\left(\bm{R}\right)\right)}\rangle_{V} }}{ \langle{ e^{\beta V\left(\bm{s}\left(\bm{R}\right)\right)}}\rangle_{V} }
    \label{eq:reweight}
\end{equation}
where $\langle{\cdot}\rangle_V$ represents a time average in the biased simulation. Another way of looking at the reweighting procedure is to rewrite Eq.~\ref{eq:reweight} as an ordinary time average in a time $t'$ scaled by the value of the bias potential:
\begin{equation}
    \begin{aligned}
    \langle{O \left(\bm{R}\right) }\rangle &= \frac{\int_{0} ^{T} dt\ O \left(\bm{R}_t\right) e^{\beta V\left(\bm{s}\left(\bm{R}_t\right)\right)} }{\int_{0} ^{T} dt\ e^{\beta V\left(\bm{s}\left(\bm{R}_t\right)\right)}} = \frac{1}{T'} \int_{0} ^{T'} dt'\ O \left(\bm{R}_{t'}\right)
    \end{aligned}
    \label{eq:time-reweight}
\end{equation}
where we have performed a change of variable  $dt' = e^{\beta V(\bm{s}(\bm{R}_t))}\ dt$, and $T'=\int_{0} ^{T} dt\ e^{\beta V\left(\bm{s}\left(\bm{R}_t\right)\right)}$ represents the total scaled time. This means that we can interpret the enhanced sampling simulation as a dynamics which samples the Boltzmann distribution on the $t'$ time scale. Thus the VAC procedure can be straightforwardly applied provided that the correlation functions of Eq.~\ref{eq:tica} are calculated in $t'$ time~\cite{McCarty2017c,Yang2018}. Additional details are reported in the materials and methods section.

It is important to note that, even though the enhanced sampling simulation in $t'$ time asymptotically samples the Boltzmann distribution, its sampling speed does differ from the unbiased one. As a result, the spectrum of the transfer operator will be different, since the degrees of freedom that have already been accelerated in the initial simulation will have a smaller contribution.
In fact, our group has previously shown that a successful enhanced sampling simulation leads to small leading eigenvalues of the transfer operator as the slow modes are accelerated~\cite{McCarty2017c}.

Among the many enhanced sampling methods that allow the reweighting procedure of eqs.~\ref{eq:reweight}-\ref{eq:time-reweight} we choose here to use OPES for a variety of reasons that will be discussed later, but first, we sketch the main features of this method.
OPES~\cite{Invernizzi2019b,Invernizzi2020UnifiedSampling} first builds an on-the-fly estimate of the equilibrium probability distribution $P(\bm{s})$ and the bias is then chosen to drive the system towards a desired target distribution: $p^{tg}(\bm{s})$.
\begin{equation}
    V\left(\bm{s}\right) = - \frac{1}{\beta} \log {\frac{p^{tg}(\bm{s})}{P(\bm{s})} }.
\end{equation}
At convergence, the free energy surface (FES) as a function of $\bm{s}$ is computed from $F(\bm{s})=-k_B T \log{P(\bm{s})}$. 
An appropriate choice of the target distribution allows one to sample a variety of ensembles including the well-tempered $P(\bm{s})^{1/\gamma}$ ~\cite{Barducci2008} with $\gamma > 1$, uniform distribution, up to generalized ensembles~\cite{Invernizzi2020UnifiedSampling}. The latter possibility is used here to sample the multithermal ensemble, where configurations relevant to a preassigned range of temperatures are sampled. This is similar to replica-exchange methods, but does not involve abrupt exchange of configurations. In the OPES version, the multithermal ensemble is sampled by using the potential energy $U(\bm{R})$ as collective variable.  

One important factor that made us choose OPES is that it reaches the quasi-static regime more rapidly than metadynamics~\cite{Invernizzi2019b}. Thus, the bias varies  more smoothly and the noise in the calculation of scaled time $t'$ correlation functions is reduced. 

\if\mycmd0
  \begin{figure*}[h!]
  \includegraphics[width=0.88\textwidth]{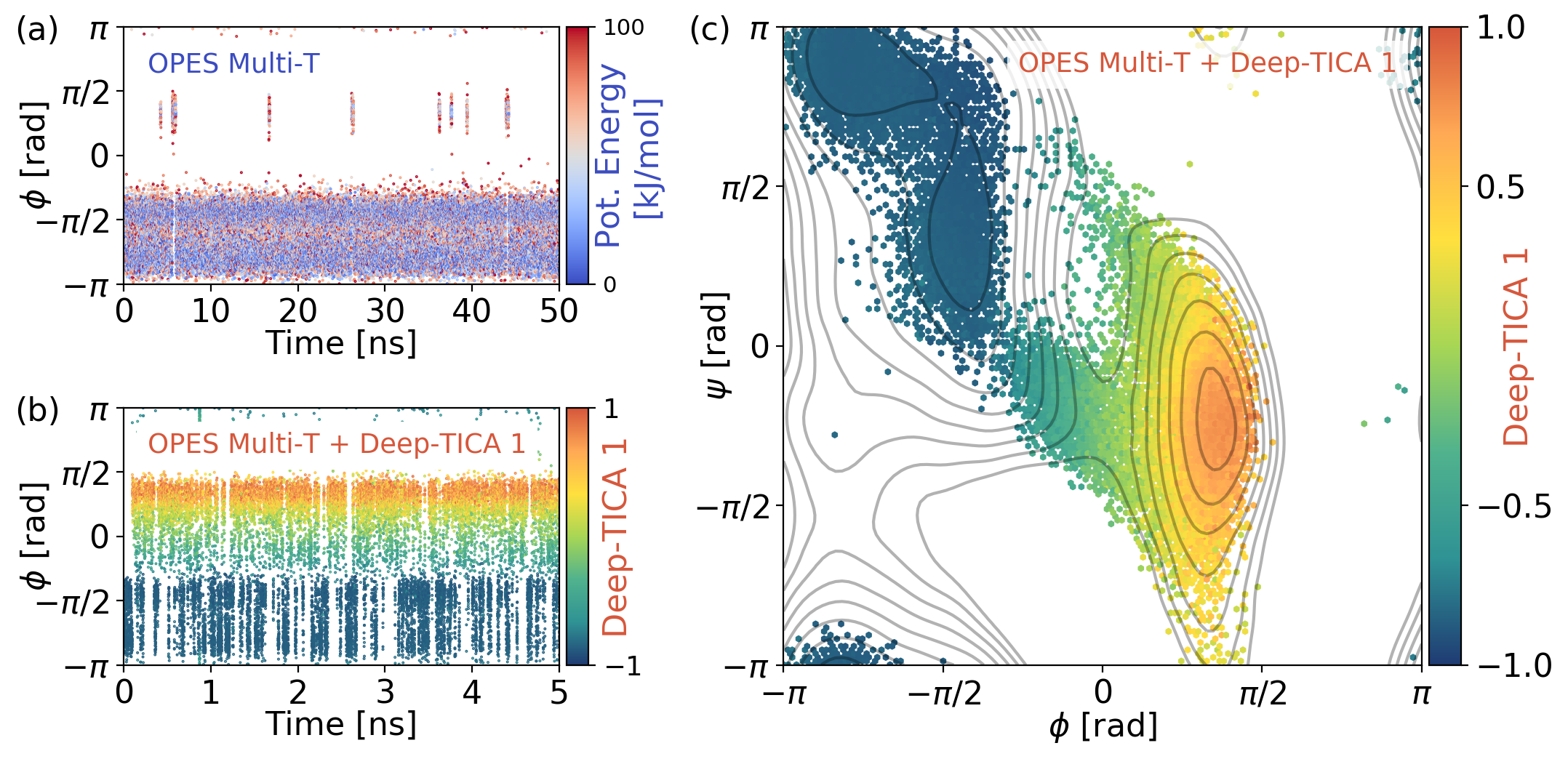}
  \else
  \begin{SCfigure*}[\sidecaptionrelwidth][t]
  \includegraphics[width=11.4cm]{figures/ala2_multi_phi_new.png}
 \fi
\centering
\caption{Deep-TICA procedure applied to a multithermal simulation of alanine dipeptide. (a) Time evolution of the $\phi$ angle in the exploratory OPES multithermal simulation, colored according to the potential energy. (b) Time evolution of the same angle for the simulation in which also the bias on Deep-TICA 1 is added, colored with the value of the latter variable. It can be seen that the system immediately reaches a diffusive behaviour. 
(c) Ramachandran plot of the configurations explored in the Deep-TICA simulation, colored with the average value of Deep-TICA 1. Grey lines denotes the isolines of the free energy surface, spaced every 2 $k_B T$. Note that the sampling is focused on the minima and the transition regions that connect them. 
}
\label{fig:ala2-multi}
\if\mycmd0
  \end{figure*}
  \else
  \end{SCfigure*}
 \fi

\subsection*{A recommended strategy}

We outline here the key steps of our recommended procedure, see also Fig.~\ref{fig:scheme}:
\begin{enumerate}[leftmargin=0.5cm]
    \item Exploration. Harness a number of reactive events using a CV-based OPES simulation with a trial CV $s_0$, multithermal sampling (in such a case  $s_0=U(\bm{R})$), or even a combination of the two. Store the final bias potential $V^*(s_0)$ of this initial simulation. 
    \item CV construction. Select the descriptors to be used as inputs of the NN.  Train the Deep-TICA CVs using the trajectories generated in step (1) by calculating the correlation functions in $t'$ time. 
    \item Sampling. Perform an OPES simulation using the leading Deep-TICA eigenfunction as CV on the Hamiltonian modified by the addition of the bias potential $V^*(s_0)$.
\end{enumerate}

This procedure can be iterated but usually at this stage the FES is well converged. In the examples presented in this work, we enhance the fluctuations of the eigenfunction associated with the slowest mode (Deep-TICA 1) while using the others for analysis purposes.

Unlike the approaches developed earlier~\cite{McCarty2017c,Yang2018}, in step (3) we also add the bias $V^*(s_0)$ to the Hamiltonian. In this way we take into account the fact that the slow modes computed in step (2) reflect the rate of convergence to the Boltzmann distribution sampled by reweighting (Eq.~\ref{eq:reweight}) from the trajectories generated using the Hamiltonian $H+V^*(s_0)$.

\section*{Results and discussion}
\subsection*{Alanine dipeptide from multicanonical simulations and CV-biased dynamics}

A simple yet informative test of enhanced sampling methods is offered by the study of the conformational equilibrium of alanine dipeptide in vacuum.  At room temperature, this small peptide exhibits two metastable states, namely the more stable $C_{7eq}$ composed of two substates and the less populated  $C_{7ax}$. The conformational transition between the two states  is well described by the torsional angles $\phi$ and $\psi$, with the former being close  to an ideal CV. However, since our scope is mostly didactical we shall on purpose ignore this information and build efficient CVs using as descriptors all the heavy atom  interatomic distances. To illustrate the flexibility and power of the method, we consider here two scenarios that differ in the way the initial reactive trajectories are generated.  

We start with illustrating the first strategy which consists of using OPES to sample the multithermal ensemble.
As can be seen from Fig.~\ref{fig:ala2-multi}a this procedure is not very efficient and promotes only a small number of transitions, and thus the free energy estimate is noisy (Fig.~\ref*{fig:si-ala2-multi-convergence}). 
Rather remarkably, this limited information is enough to extract the slow modes of the system using Deep TICA and obtain CVs that are efficient in promoting sampling. We find the training to be robust concerning the choice of lag-time (Fig.~\ref*{fig:si-ala2-training-lagtime}), and also to the number of configurations used (Fig.~\ref*{fig:si-ala2-training-configs}). The leading Deep-TICA 1 variable is associated with the transition between $C_{7eq}$ and $C_{7ax}$, while the second describes the transition between the $C_{7eq}$ substates (Fig.~\ref*{fig:si-ala2-isolines}).

Subsequently, a new OPES multithermal simulation is performed biasing in addition also Deep-TICA 1. 
The first remarkable result is a two hundred-fold increase in the number of transitions per unit time as compared to the initial simulation (Fig.~\ref{fig:ala2-multi}b). The system immediately reaches a quasi-static regime in which the average interval between inter-state transitions is about 25ps, to be compared to the pure multithermal simulation in which the same rate was as large as 3 ns.  As a consequence of this speed-up, the free energy difference between metastable states converges in just 1 ns to the reference value within 0.1 $k_B T$. A quantitative analysis of the convergence of the simulations can be found in the supporting information (SI, Fig.~\ref*{fig:si-ala2-multi-convergence}), together with a comparison of the simulations with and without the static bias potential $V^*(s_0)$ and the free energy surface as a function of the two Deep-TICA CVs (Fig.~\ref*{fig:si-ala2-fes}).

The time to convergence is comparable to what one finds when using as CVs the physically informed dihedral angles $\phi$ and $\psi$~\cite{Invernizzi2019b}, \emph{but} here it is the result of a procedure that is generally applicable and  does not require any previous understanding of the system. There is also a clear improvement with respect to discriminant-based CVs that use the same set of  descriptors~\cite{Bonati2020Data-DrivenSampling}.

Notably, the Deep-TICA CV promotes sampling along the two different pathways connecting $C_{7eq}$ and $C_{7ax}$ (Fig.~\ref{fig:ala2-multi}c), as it measures the sampling progress along all transition pathways~\cite{McGibbon2017IdentificationDynamics}. Combined with the OPES ability to set an upper bound to the value of the added bias, this results in focusing the sampling on the most interesting parts of the FES which are the minima and the transition region. 
\begin{figure}[h!]
\includegraphics[width=0.99\linewidth]{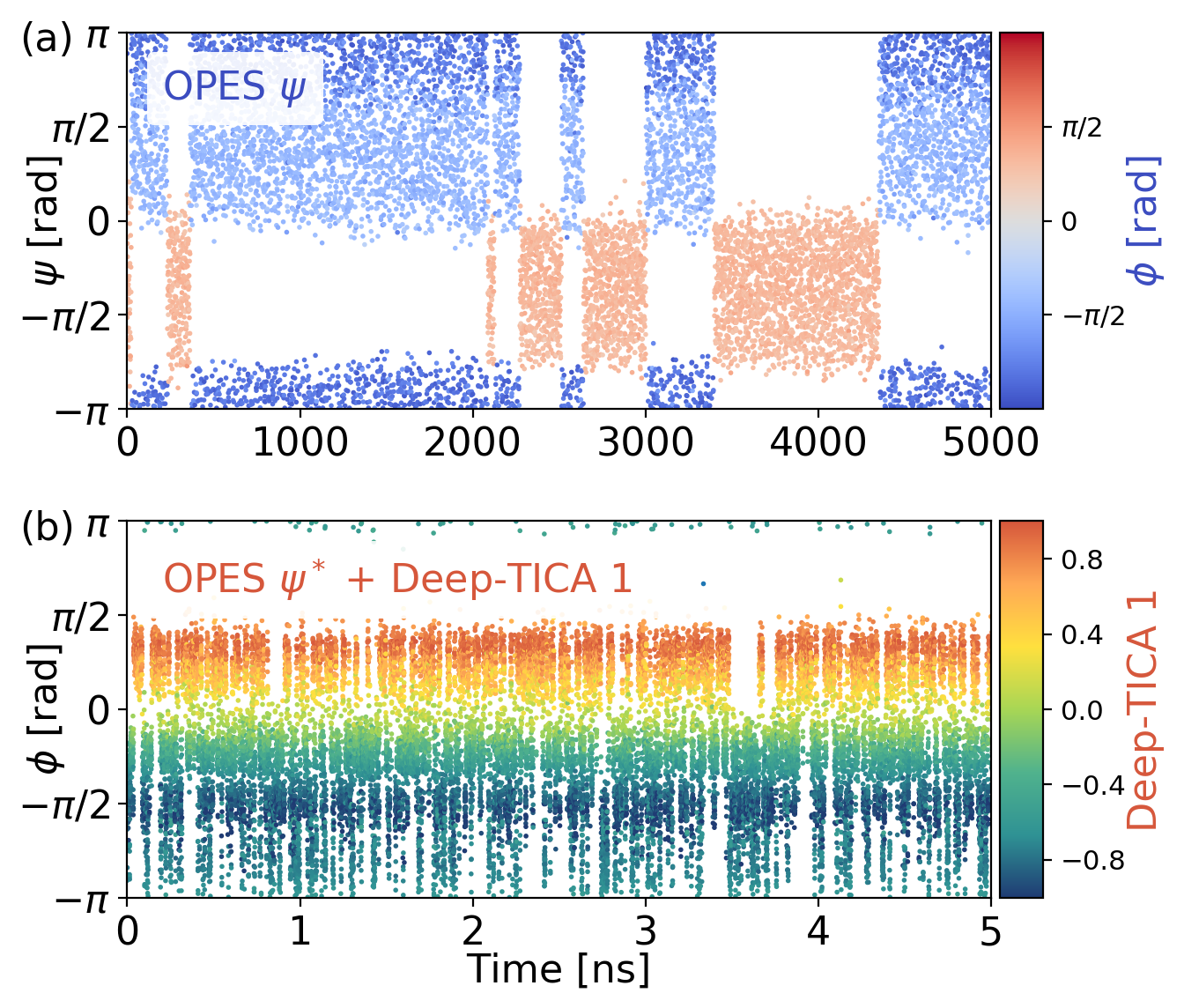}
\centering
\caption{(a) Time evolution of the $\psi$ angle in the exploratory simulation driven by $\psi$. The points are  colored with the values of the $\phi$ angle. (b) Time evolution of the $\phi$ angle in the final Deep-TICA simulation, colored with the value of Deep-TICA 1. This results in a diffusive simulation similar to the previous example, which is even more impressive here given the poor quality of the exploratory sampling.}
\label{fig:ala2-psi}
\end{figure}

The above-described procedure was based on the ability of the initial multithermal simulation to induce transitions between the local minima.  However, in many cases multithermal simulations are not able to induce even a single transition and the use of a CV to generate a reactive trajectory is called for. Thus we found instructive to exemplify the performance of Deep-TICA when the initial biased simulation is driven by a CV. Again we want to challenge the method and we choose the angle $\psi$ as starting CV. A cursory look at the alanine dipetide FES of Fig.~\ref{fig:ala2-multi}c makes one realize that $\psi$ is a very poor CV being almost perpendicular to the direction of the most likely transition paths. For this reason, it is an exemplary case of a CV that should not be used~\cite{Invernizzi2019}. 
The low quality of the CV is reflected in the fact that we need to simulate the system for 5 microseconds to observe a handful of transitions (Fig.~\ref{fig:ala2-psi}a). As before, we feed these  scant data to the Deep-TICA machinery and compute the highest eigenfunctions of the transfer operator. When we perform a new OPES calculation biasing Deep-TICA 1, the simulation immediately reaches a diffusive regime similarly to the previous example (Fig.~\ref{fig:ala2-psi}b), which allows converging the free energy in a very short timescale of 1 ns (Fig.~\ref*{fig:si-ala2-psi-convergence}).
This is possibly an extreme example but shows that remarkable speedups can be attained when the slow modes are correctly identified and their sampling accelerated. In real life, one tries not to use CVs as bad as $\psi$ but the use of suboptimal CVs is far from rare. In this respect, the Deep-TICA method holds the promise of remedying  a  poor initial CV choice.

As discussed in the previous sections, the eigenfunctions that we obtain describe the slowly converging modes of the sampling dynamics performed at step (1). In the SI we investigated this point by comparing the CVs extracted from the multicanonical simulation and those obtained from the $\psi$-biased dynamics, highlighting the effect of initial simulation (Fig.~\ref*{fig:si-ala2-isolines}). Furthermore, it should be noted that for these cases in which the quality of the initial CV $s_0$ is poor the advantage of using the bias $V^*(s_0)$ from the previous simulation is less significant but still non-negligible (see Figs.~\ref*{fig:si-ala2-multi-convergence}-\ref*{fig:si-ala2-psi-convergence}).

\subsection*{A blind approach to chignolin folding}

Chignolin is one of the smallest proteins that can be folded into a stable structure. Here we focus on its variant CLN025, which has been extensively studied using molecular simulations by performing long simulations on the Anton supercomputer~\cite{Lindorff-Larsen2011HowFold} and
using enhanced sampling techniques~\cite{Okumura2012TemperatureMethod,Shaffer2016a,McKiernan2017ModelingFormation,Mendels2018a}. 

\begin{figure}[h!]
\includegraphics[width=\linewidth]{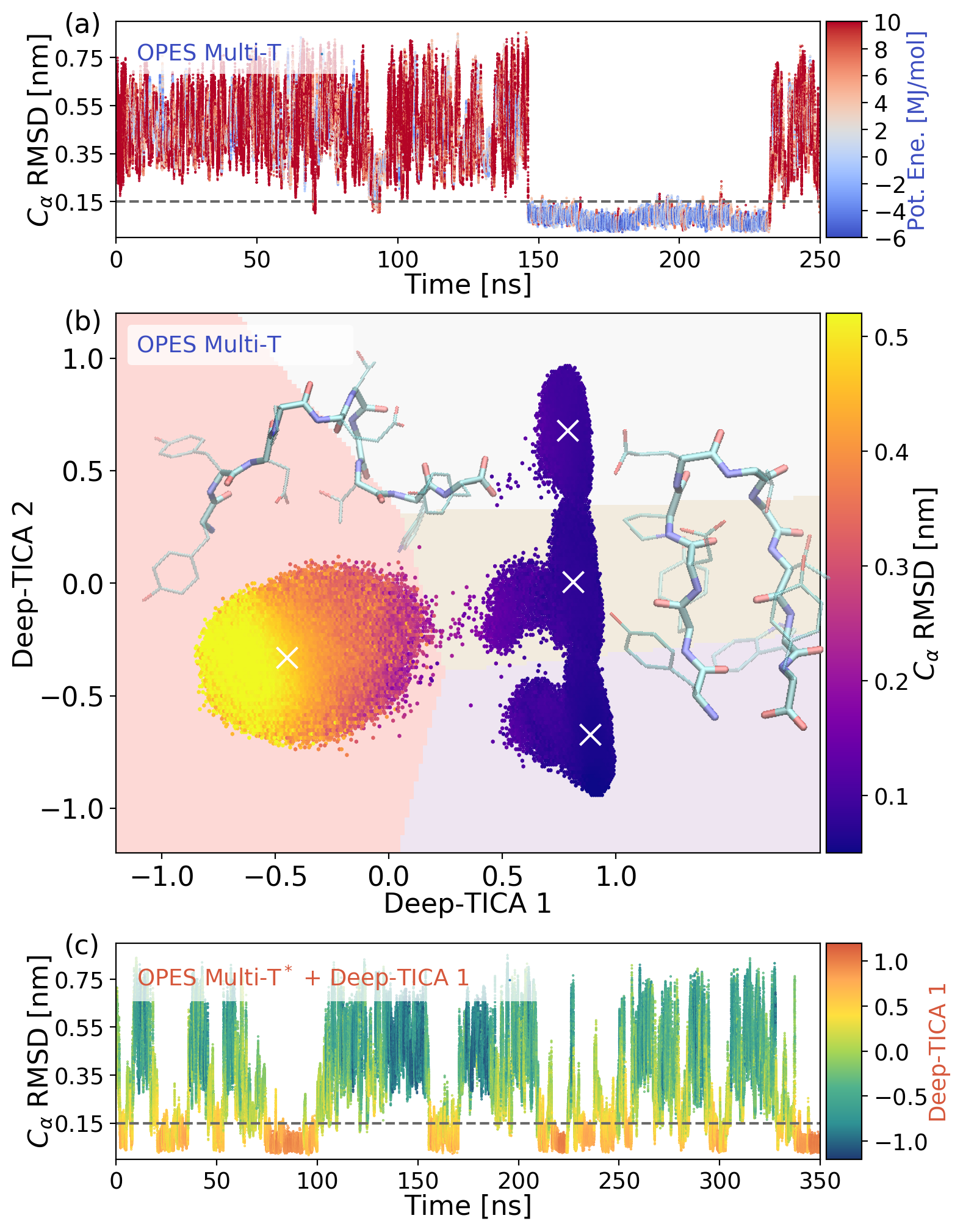}
\centering
\caption{Deep-TICA procedure applied to chignolin folding. (a) Time evolution of the $C_\alpha$ RMSD for one replica during the initial multithermal run. The points are colored according to their potential energy value. Low energy values reflect the fact that configurations relevant at lower temperatures are sampled.
(b) Scatter plot of the two leading Deep-TICA CVs in the exploratory simulation. Points are colored according to the average $C_\alpha$ RMSD values. A weighted k-means clustering identifies four clusters whose centers are denoted by a white X. The pale background colors reflect how space is partitioned by the clustering algorithm. Snapshot of chignolin in the folded (high values of Deep-TICA 1) and unfolded (low values) states are also shown, realized with VMD~\cite{Humphrey1996VMD:Dynamics}. 
(c) Time evolution of $C_\alpha$ RMSD for a replica in the multithermal simulation biasing also Deep-TICA 1, colored with the value of the latter variable. Time evolution for the other replicas is reported in Fig.~\ref*{fig:si-chignolin-traj}.}
\label{fig:chignolin-traj}
\end{figure}

Once again we pretend that we are unaware of the progress made in the understanding of Chignolin behavior and follow the same blind approach pursued in the first alanine dipeptide simulation reported above. That is, in the exploratory phase we perform an OPES multithermal sampling, this time boosted by the use of multiple replicas (Fig.~\ref{fig:chignolin-traj}a). This leads to observing a few folding-unfolding events. Using these trajectories we construct a Deep TICA CV using as descriptors all the 4278 interatomic distances between heavy atoms. 

Since enhancing the sampling of a CV that depends on thousands of descriptors would have been computationally inefficient, we decided to reduce their number by selecting the most relevant one for the leading CV via a sensitivity analysis~\cite{Bonati2020Data-DrivenSampling}. In this way, we selected 210 descriptors (which are reported also in Fig.~\ref{fig:scheme}). We retrain the NN using this reduced set and find out that the leading eigenvalue is only decreased by just 0.5 \%, thanks to the variational flexibility of the NN. Interestingly, the selected distances involve both backbone and side chain atoms, suggesting that also the latter have a significant role in the folding process.

As expected, the first CV (Deep-TICA 1) describes the folded to unfolded transition that is the slowest mode of the system.  Instead, the second one (Deep-TICA 2) characterizes the fine structure of the folded state, as we will discuss later. This can be seen in  Fig.~\ref{fig:chignolin-traj}b, where we colored the points sampled in the initial trajectory with the value of the backbone $C_{\alpha}$ root-mean-square deviation (RMSD). It should be noted that we find no evidence of stable misfolded states along the dominant folding collective variables, in agreement with simulations using the same force-field ~\cite{Lindorff-Larsen2011HowFold,Kuhrova2012Force-fieldRedesign,McKiernan2017ModelingFormation}.

Performing a new simulation with a bias potential along Deep-TICA 1 results in an enhanced sampling of the transition region (Fig.~\ref*{fig:si-chignolin-ene-tica-scatter}), with a 20-fold increase in the rate of folding events compared to the multithermal simulation (Fig.~\ref{fig:chignolin-traj}c). Due to the use of the multithermal approach all the free energy profiles in the chosen range of temperatures can be calculated with great accuracy (Fig.~\ref*{fig:si-chignolin-fes2d-1-temp}-\ref*{fig:si-chignolin-fes1d-temp}). Indeed, the average statistical error calculated with a weighted block-average technique~\cite{Invernizzi2020UnifiedSampling} is about 0.5 kJ/mol, improving on classifier-based approaches applied to the same system~\cite{Mendels2018a}. In particular, we find an excellent agreement with an unbiased 106 $\mu$s reference trajectory~\cite{Lindorff-Larsen2011HowFold} at $T=340K$ (Fig.~\ref*{fig:si-chignolin-fes-comparison}). Note that a study of the protein behavior at lower temperatures using standard MD would have been significantly more difficult to perform.

\begin{figure}[h!]
\includegraphics[width=\linewidth]{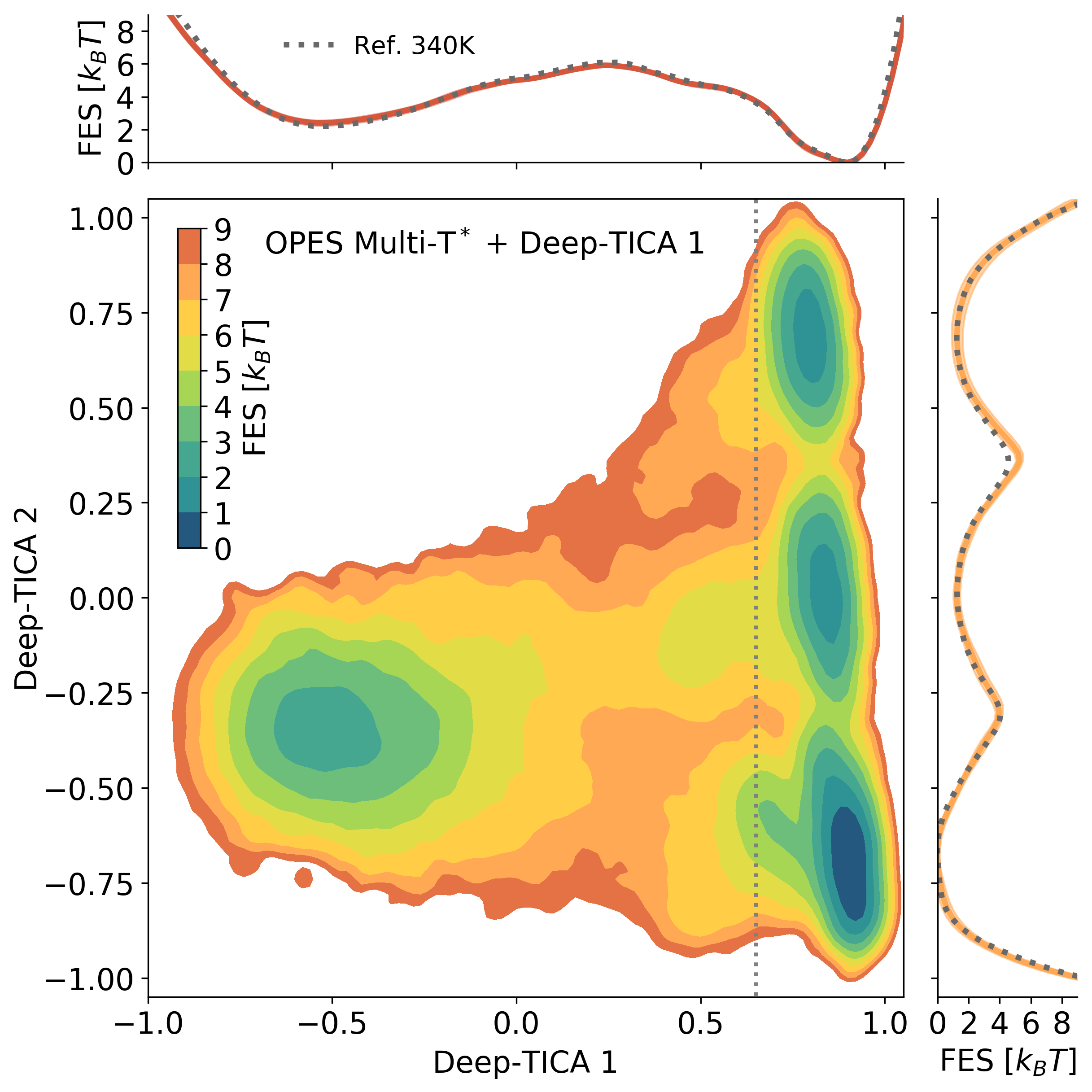}
\centering
\caption{Free energy surface of chignolin at T=340K as a function of the two leading Deep-TICA CVs. Above and to the right are shown the projections of the FES along the corresponding axis (solid line), confronted with the reference value obtained from a long unbiased MD trajectory at 340K~\cite{Lindorff-Larsen2011HowFold} (dotted line). Note that the projection of Deep-TICA 2 is obtained by integrating only the region of space with Deep-TICA 1 > 0.65 (marked by a dotted line in the central panel), to highlight the barriers between the folded metastable states. }
\label{fig:chignolin-fes}
\end{figure}

In Fig.~\ref{fig:chignolin-fes} we report the FES relative to $T=340K$ plotted as a function of the two leading eigenfunctions, together with their projections along with Deep-TICA CVs. We find two major states divided by a significant free energy barrier, which correspond to the unfolded and folded basins. However, the latter exhibits a fine substructure that can be traced back to the Threonine sidechains (THR6 and THR8) occupying different dihedral conformations (Fig.~\ref*{fig:si-chignolin-sidechain}). At 340K these states can interconvert on the timescale of nanoseconds, but, of course, this time is significantly slower at lower temperatures (see Fig.~\ref*{fig:si-chignolin-fes2d-2-temp}). Notably, the most likely conformation is stabilized by the presence of a hydrogen bond between the two THR sidechains (Table~\ref*{tab:hbonds}), as previously observed in a structural analysis study for wild-type chignolin~\cite{Maruyama2018AnalysisChignolin}. This is remarkable because it was discovered without any prior knowledge of neither structural conformation of the system nor the dynamics of folding, and it suggests that going beyond backbone-only structural descriptors is necessary to obtain an accurate representation of the folding dynamics.

\subsection*{Improved data-driven description of silicon crystallization}

Silicon crystallization is a first-order phase transition hindered by a large free energy barrier. 
This implies that in step (1) of the Deep-TICA procedure one has to resort to CV-based simulations to harness reactive trajectories.  Our group has previously investigated the application of TICA to simulate Na and Al crystallization~\cite{Zhang2019}. The study of Si crystallization is however more difficult, due to the directional nature of the bonds and the ease with which defective and glassy structures can arise.

To address this problem, we make use of a recently developed set of descriptors that have proven to be useful in a machine-learning context~\cite{Karmakar2021CollectiveCrystallisation}. These are the peaks of the three-dimensional structure factor of a crystal that is commensurate with the MD simulation box. Compared with spherically averaged structure factors used in Ref.~\cite{Bonati2018,Zhang2019}, these descriptors have the advantage that they facilitate the formation of crystal structures aligned with the axes of the box. Since they measure the presence of long-range order in the system they are a natural choice in the study of crystallization. 

In Ref.~\cite{Karmakar2021CollectiveCrystallisation} these peaks have been combined into a CV using the Deep-LDA classification method. The question that we address here is whether we can improve upon the Deep-LDA description and obtain a CV that incorporates dynamical information. 
In order to make a fair comparison between Deep-LDA and Deep-TICA we use the same set of descriptors chosen with a well-defined universal concept. That is, we use in both cases the first 95 $S(\bm{k})$ Bragg peaks with modulus $k \leq 7\ A^{-1}$ (Fig.~\ref*{fig:si-silicon-peaks}), which amounts as fixing the CV spatial resolution. 
 
\begin{figure}[t!]
\includegraphics[width=\linewidth]{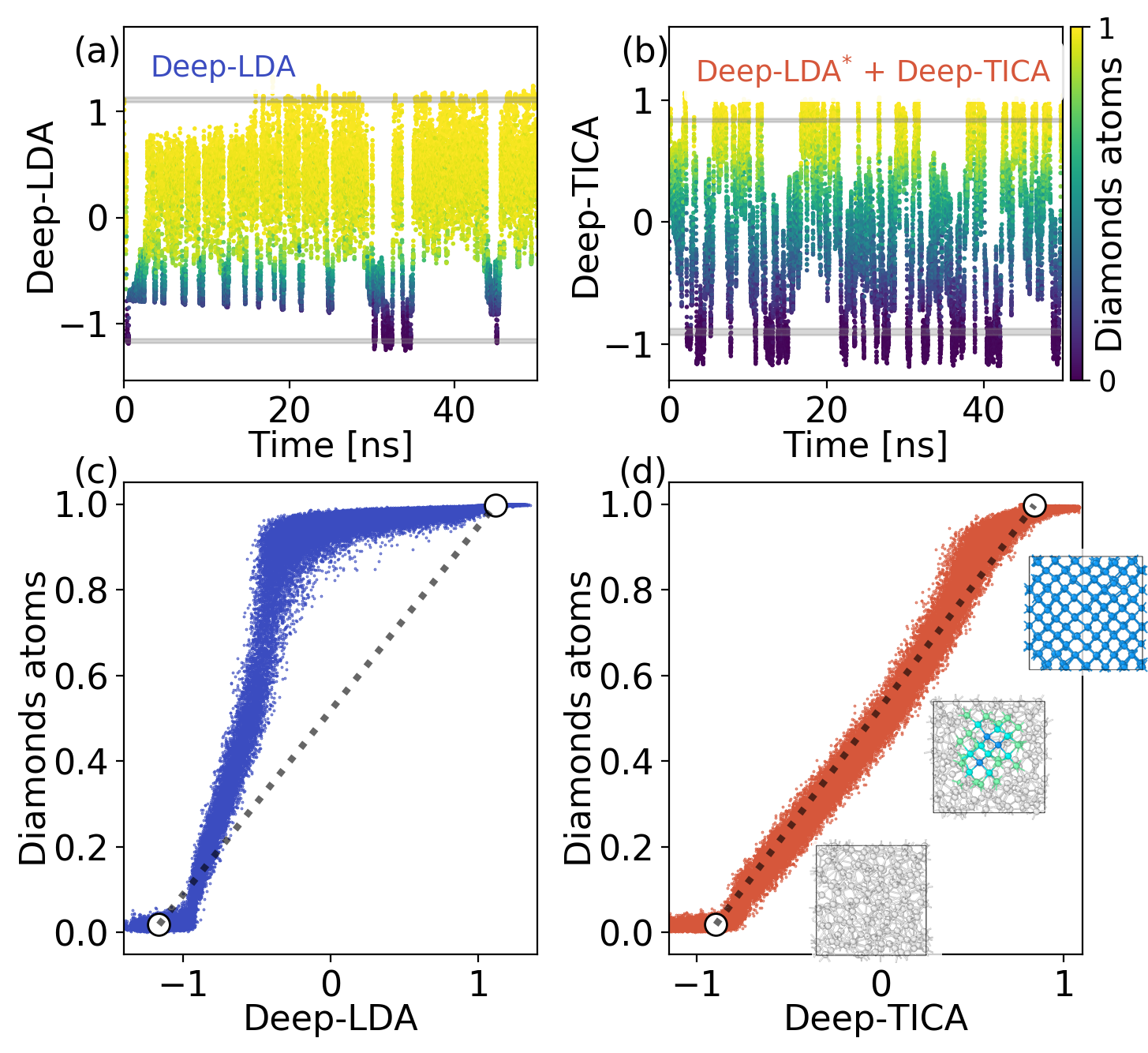}
\centering
\caption{Comparison between a Deep-LDA driven simulation (left) and the one based on the present Deep-TICA approach (right). In the top row we report the time evolution of (a) Deep-LDA CV in the initial simulation and of (b) Deep-TICA CV in the improved one. In both panels the points are colored according to the fraction of diamond-like atoms in the system, computed as in Ref.~\cite{Piaggi2019CalculationEnsemble}. Grey shaded lines indicate the values of the two CVs in unbiased simulations of the liquid (bottom lines) and solid (top lines). Panels (c) and (d) report the correlation between the two data-driven CVs and the fraction of diamond-like atoms. White circles denote the mean values of the two CVs in the liquid and solid states, while the dotted grey line interpolates between them. In panel (d) we report also a few snapshots of the crystallization process made with OVITO~\cite{Stukowski2010}.}
\label{fig:silicon}
\end{figure}

The Deep-LDA CV is trained using short MD simulations in the liquid and the cubic diamond states. Afterward, an OPES simulation is performed, which promotes a few crystallization and melting events, though the system struggles to find its way to the liquid state (Fig.~\ref*{fig:silicon}a). From this trajectory a Deep-TICA CV is extracted and subsequently used to enhance sampling together with the final static bias $V^*(s_0)$ (Fig.~\ref*{fig:silicon}b). Similar to what happened in the previous examples, this procedure leads to an increase in the number of transitions between the solid and liquid states, which allows converging the free energy estimate already after 20 ns. The statistical uncertainty on the free energy difference is reduced compared to the Deep-LDA simulation (Fig.~\ref*{fig:si-silicon-convergence}).
Furthermore, the free energy difference between the two states is close to zero at T=1700K, in excellent agreement with the melting point of the interatomic potential~\cite{Stillinger1985,Bonati2018}.

As a final comment, we argue that the different degree of sampling efficiency between the two data-driven CVs has to be rooted in their different training objectives. Being trained as a classifier, Deep-LDA very accurately discriminates between the solid and the liquid phase, but it has no information about the transition region which connects them. Consequently, in almost the entire range of values spanned by the variable during the simulation, the system is either completely in the crystalline or the liquid phase (Fig.~\ref*{fig:silicon}c).  
The Deep-TICA CV, besides classifying the states, reflects also the transition dynamics. In fact, in Fig.~\ref*{fig:silicon}d we see that it describes more smoothly the transition between the two phases, as  Deep-TICA is linearly correlated with the number of crystalline atoms, which is a relevant quantity in the classical nucleation theory framework~\cite{Kelton2010NucleationBiology}.

\section*{Conclusions}

The extension of the variational principle of conformation dynamics to enhanced sampling data~\cite{McCarty2017c} represents a promising way to address the chicken-and-egg dilemma intrinsic to the determination of collective variables. Here, we leverage the flexibility of neural networks and recent developments in advanced sampling techniques to construct a general and robust protocol.  
The Deep-TICA method allows us to analyze a biased simulation trajectory, extract the slow modes which hinder its convergence, and subsequently accelerate them. This can be used to extract CVs from generalized ensemble simulations and to complement approximate CVs constructed based on physical considerations or in a data-driven manner. Besides improving sampling, this method provides us with atomistic details on the rare events dynamics.
Remarkably, our work underlines the fact that even a partial information about the transition pathways does go a long way to solve the rare event problem.
In fact, the test on the alanine dipeptide benchmark shows that the procedure is applicable even when starting from a very poor initial enhanced sampling simulation. This promises to be of great help in the study of realistic systems, where the identification of appropriate CVs is challenging.
Application of the method to the more complex examples of chignolin folding and material crystallization illustrates how this acceleration allows the FES to be reconstructed with high accuracy, with no need of physical or chemical insight into the transition dynamics.
We are confident that our approach can be applied to even more complex systems and that it can be of great help to the broad molecular simulation community.

\section*{Materials and methods}
\if\mycmd0
  \textbf{Time-lagged covariance matrices}$\quad$ {\small
  \else
  \subsection*{Time-lagged covariance matrices}
\fi
Given an enhanced sampling simulation we first rescale the time according to Eq.~\ref{eq:time-reweight}. We then search for pairs of  configurations distant a lag-time $\tau$ in time $t'$. Due to time reweighting, the value of $\tau$ cannot be interpreted as a physical time. However, we found consistent results for a range of lag-time values such that all desired eigenvalues did not decay to zero (Fig.~\ref{fig:si-ala2-training-lagtime}). Note that, when Eq.~\ref{eq:time-reweight} is discretized, time intervals become unevenly spaced in $t'$ and the calculation of the time-lagged covariance matrices requires some care. To deal with this numerical issue we resort to the procedure proposed in \cite{Yang2018}. These pairs of configurations are saved in a dataset and later used for the NN training. Furthermore, it should be noted that while in principle the two correlation matrices are symmetric, this condition might not be satisfied when estimating them from MD simulations due to limited sampling. Here we symmetrize the matrices to enforce detailed balance as $C_{ij} ^{sym} = (C_{ij}+C_{ji})/2$. This choice is the simplest, although it introduces a bias~\cite{Wu2017VariationalSimulations}. 

\if\mycmd0
  }\textbf{Deep-TICA CVs training}$\quad${\small
  \else
  \subsection*{Deep-TICA CVs training}
\fi
Deep-TICA CVs are trained using the machine learning library PyTorch~\cite{AdamPaszkeSamGrossetal2017}. As previously done for Deep-LDA and other non-linear VAC methods~\cite{Chen2019}, we apply Cholesky decomposition to $C(0)$ to convert Eq.~\ref{eq:tica} into a standard eigenvalue problem. This allows to back-propagate the gradients through the eigenvalue problem by using the automatic differentiation feature of the  ML libraries. We use a feed-forward neural network composed by 2 layers and the hyperbolic tangent as activation function. The NN parameters is optimized using ADAM with a learning rate of 1e-3. To avoid overfitting, we split the dataset in training/validation and apply early stopping with a patience of 10 epochs. Furthermore, the inputs are scaled to have zero mean and variance equal to one. Also the Deep-TICA CVs are scaled in order for their range of values to be between -1 and 1. The normalization factors are calculated over the training set and saved into the model for inference. Once the training is performed, the model is serialized so that it can be used on-the-fly in a molecular dynamics simulation. 

\if\mycmd0
  }\textbf{PLUMED-Pytorch interface}$\quad${\small
  \else
  \subsection*{PLUMED-Pytorch interface}
\fi
To use the transfer operator eigenfunctions as CVs for enhanced sampling simulations, we use a modified version of the open-source PLUMED2~\cite{Tribello2014} plug-in which we interfaced with the LibTorch C++ library, as in Ref.~\cite{Bonati2020Data-DrivenSampling}. The model trained in Python is loaded by PLUMED to evaluate CV values and derivatives with respect to descriptors for new configurations explored during the simulation and apply a bias potential along them.
Input files to run the simulations will be deposited in the PLUMED-NEST~\cite{Bonomi2019} repository. 

\if\mycmd0
  }\textbf{Alanine dipeptide simulations}$\quad${\small
  \else
  \subsection*{Alanine dipeptide simulations}
\fi
Alanine dipeptide (ACE-ALA-NME) simulations are carried out using GROMACS~\cite{VanDerSpoel2005a} patched with PLUMED. We use the Amber99-SB~\cite{Hornak2006} force field with a time step of 2 fs. The NVT ensemble is sampled using the velocity rescaling thermostat~\cite{Bussi2007} with a temperature of 300K. 
For the OPES multithermal simulation we sample a range of temperatures from 300K to 600K, updating the bias every PACE=500 steps. We run a 50 ns simulation, and use the last 35 ns where the bias is in a quasi static regime. We then look for configurations separated by a lag time of 0.1. The input descriptors of the Deep-TICA CVs are the 45 distances between the heavy atoms, and the following NN architecture is used: 45-30(tanh)-30(tanh)-3. We optimize the first 2 eigenvalues in the loss function. After training the CVs, a new OPES simulation is performed in the multi-thermal ensemble with the same parameters as before, combined with the multi-umbrellas OPES ensemble along  Deep-TICA 1 CV with the parameters SIGMA=0.1 and BARRIER=40. Since in the simulation driven by Deep-TICA 1 the time needed to converge the multithermal bias is very short, we did not use the static static bias from the previous simulation but we optimized it together with the bias along the TICA CV.

The second example involves the OPES simulation in which the dihedral angle $\psi$ is used as the CV. The parameters of OPES are PACE=500, SIGMA=0.15 and BARRIER=40. The first 500 ns out of a total simulation length of 5 $\mu$s are discarded while the remaining are used to compute time correlation functions with a lag-time equal to 5. The NN details are the same as in the multithermal example. Next, an OPES simulation is performed using Deep-TICA 1 as the CV, with parameters PACE=500, SIGMA=0.025 and BARRIER=30, along with the static bias $V^*(s)$ from the previous simulation.

\if\mycmd0
  }\textbf{Chignolin simulations}$\quad${\small
  \else
  \subsection*{Chignolin simulations}
\fi
Simulations of the CLN025 peptide (sequence TYR-TYR-ASP-PRO-GLU-THR-GLY-THR-TRP-TYR) are performed using GROMACS patched with PLUMED. Computational setup is chosen to make a direct comparison with Ref.~\cite{Lindorff-Larsen2011HowFold}. CHARMM22* force field \cite{Piana2011HowParameterization} and TIP3P water model~\cite{Jorgensen1983ComparisonWater} are used, the integration timestep is 2 fs, and the target temperature of the thermostat is set to 340K. ASP,GLU residues as well as the N- and C-terminal amino acids are simulated in their charged states. Simulation box contains 1906 water molecules, together with two sodium ions that neutralize the system. The linear constraint solver (LINCS) algorithm is applied to every bond involving H atoms and electrostatic interactions are computed via the particle mesh Ewald scheme, with a cutoff of 1 nm for all non-bonded interactions.

The initial simulation is performed with OPES to simulate the multithermal ensemble in a range of temperatures from 270K to 700K. We simulate 8 replicas sharing the same bias potential to harvest more transitions. The simulation time is 250 ns, of which the first 50 ns are not used for the Deep-TICA training. A lag-time equal to 5 is used. In order to compute the scaled-time correlation functions we reweight at the simulation temperature of 340K. Note that when starting from a multithermal simulation, one could reweight also at different temperatures and extract the associated eigenfunctions. Initially, a larger NN is trained using as input all the heavy atoms distances, with an architecture 4278-256(tanh)-256(tanh)-5. After performing an analysis of the features' relevance, based on the derivatives of the leading eigenfunction with respect to the inputs~\cite{Bonati2020Data-DrivenSampling}, a smaller NN is trained using a reduced set of 210 distances and architecture 210-50(tanh)-50(tanh)-5. The list of the distances used is available in the PLUMED-NEST repository. We observe very similar results in terms of the extracted eigenvalue when using between 100 and 300 inputs. 
The number of optimized eigenvalues in the loss function is equal to 2. Finally, we enhance the fluctuations of the Deep-TICA 1 CV via an OPES simulation with parameters PACE=500, SIGMA=0.1 and BARRIER=30, together with the static multithermal potential from the initial simulation.

\if\mycmd0
  }\textbf{Silicon simulations}$\quad${\small
  \else
  \subsection*{Silicon simulations}
\fi
Silicon simulations are carried out using LAMMPS~\cite{Plimpton95} patched with PLUMED, using the Stillinger-Weber interatomic potential~\cite{Stillinger1985}. A 3x3x3 supercell (216 atoms) is simulated in the NPT ensemble with a timestep of 2 fs. A thermostat with a target temperature of 1700K is used with a relaxation time of 100 fs, while the values for the barostat are 1 atm and 1 ps. 

First, two 5 ns long simulations of standard MD in the solid and liquid states are performed. The values of the 95 three-dimensional structure factor peaks in these configurations are computed and this information is used to construct a Deep-LDA CV, using a two-layer NN with 30 nodes per layer. A 50 ns OPES simulation biasing this variable, with PACE=500, adaptive sigma, and BARRIER=1000 is performed. The first 25 ns are not used for NN training. A lag-time of 0.5 is used. The input descriptors and architecture of the NN are the same as those used for Deep-LDA. Only the principal eigenvalue is optimized in the loss function. We then run a new OPES simulation biasing the Deep-TICA CV using the same parameters as in the initial simulation, along with the static bias potential $V^*(s_0)$. The fraction of diamond-like atoms is computed in PLUMED with the Environment Similarity CV, with parameters SIGMA=0.4 LATTICE\_CONSTANTS=5.43 MORE\_THAN=\{R\_0=0.5 NN=12 MM=24\}.

\if\mycmd0
  }
\fi

\subsection*{Acknowledgments}
{\small
The authors are grateful to Dr. Michele Invernizzi for several valuable discussions and to him and Dr. Andrea Rizzi for carefully reading the paper.
The calculations were carried out on the Euler cluster of ETH Zurich.

\vspace{1em}
Contact details: $^*$ luigi.bonati@iit.it; 
$^\dagger$ michele.parrinello@iit.it.
}

\AtNextBibliography{\footnotesize}
\printbibliography

\clearpage

\renewcommand{\thesection}{S\arabic{section}}
\renewcommand{\thesubsection}{S\arabic{subsection}}
\renewcommand{\thefigure}{S\arabic{figure}}
\renewcommand{\theequation}{S\arabic{equation}}
\renewcommand{\thetable}{S\arabic{table}}
\renewcommand{\thepage}{SI-\arabic{page}}
\setcounter{page}{1}  
\setcounter{figure}{0}  
\setcounter{equation}{0}  

\onecolumn
\section*{SUPPORTING INFORMATION}
\if\mycmd1
  ~
 \fi
\section*{A. ALANINE DIPEPTIDE SIMULATIONS}
~
\subsection*{Deep-TICA training}
~
\vfill
\begin{figure}[h!]
\if\mycmd0
  \captionsetup{font={normalsize},textfont={normalfont}}
\fi
\includegraphics[width=\linewidth]{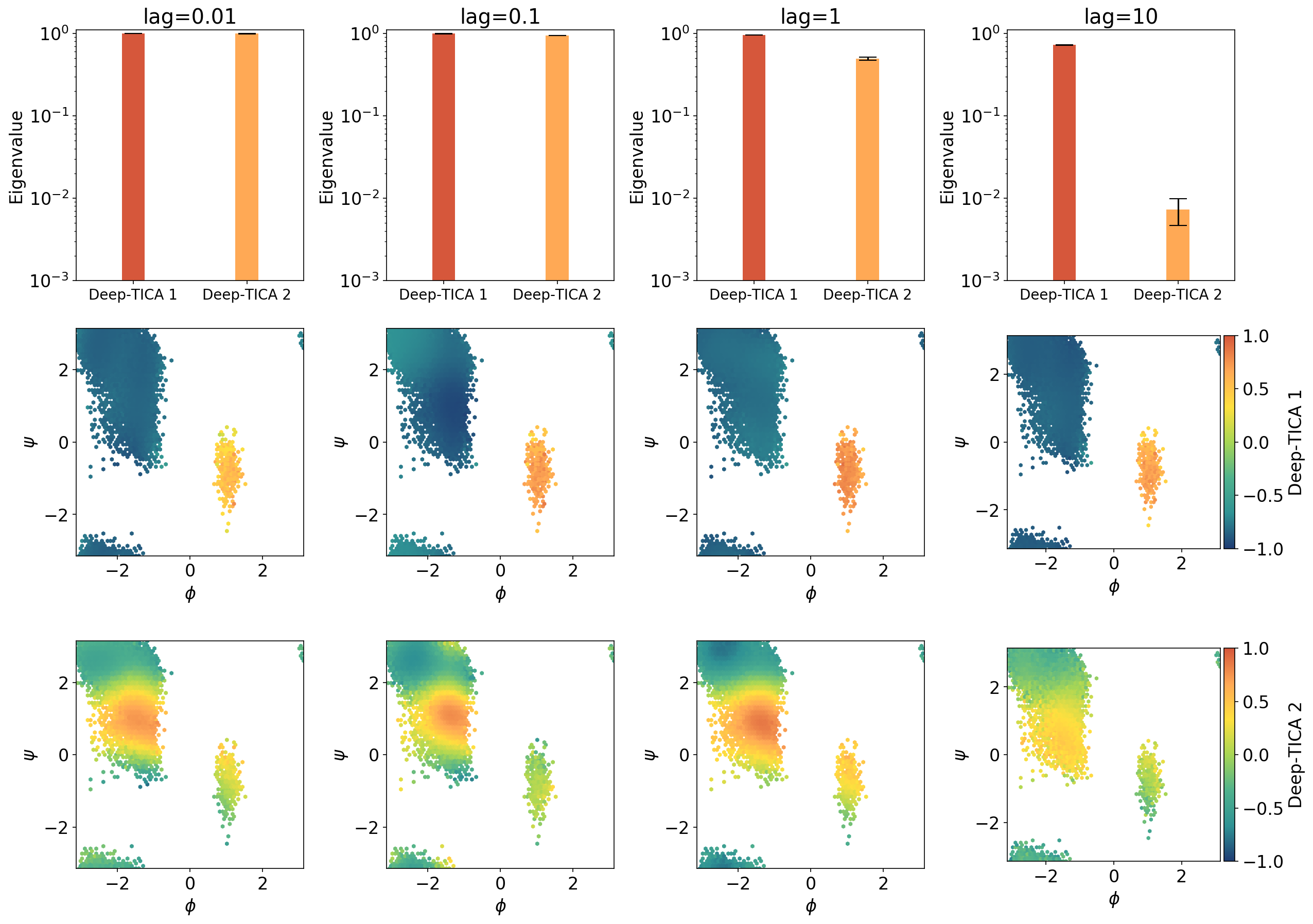}
\centering
\caption{\textbf{Deep-TICA CVs training versus lag-time} for the alanine dipeptide multithermal example. Columns correspond to different values of the lag-time used. Top row: average value of the two leading eigenvalues (bars) and standard deviation over 5 repeated trainings (black lines). Mid and bottom rows: training points projected on the $\phi-\psi$ plane, colored according to the related eigenfunctions Deep-TICA 1 and 2. The first eigenfunction is always consistent across the range of lag times studied, while the second eigenfunctions starts to lose signal when the associated eigenvalue becomes too small. This suggests to choose the value of the lag-time such that all the desired eigenvalues have not decayed to zero.}
\label{fig:si-ala2-training-lagtime}
\end{figure}
\vfill

\clearpage

\vfill
\begin{figure}[p]
\if\mycmd0
  \captionsetup{font={normalsize},textfont={normalfont}}
\fi
\includegraphics[width=\linewidth]{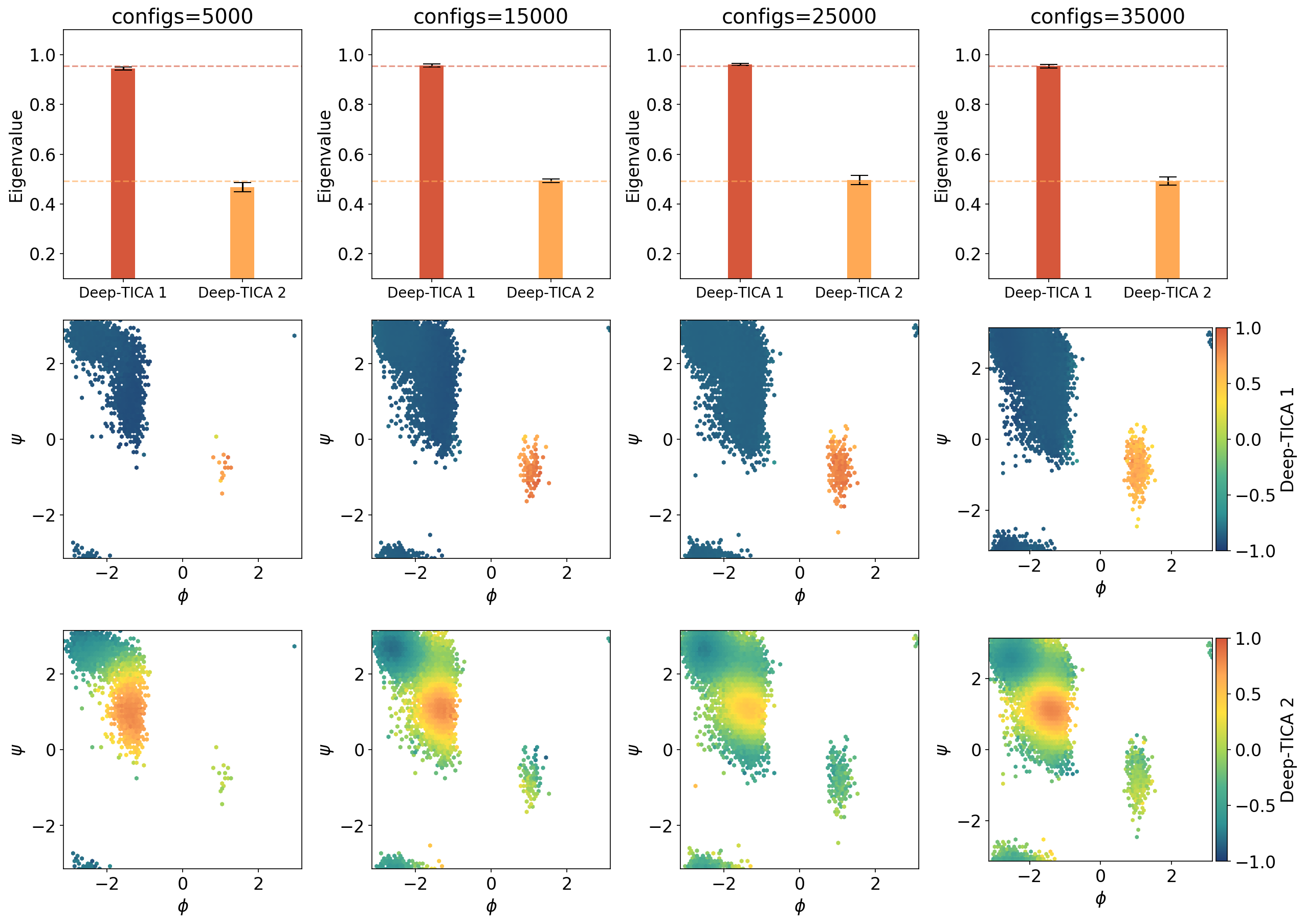}
\centering
\caption{\textbf{Deep-TICA CVs training versus number of configurations} used for the training for the alanine dipeptide multithermal example. The configurations are extracted every 1 ps after 15 ns. Columns correspond to different values of the lag-time used. Top row: average value of the two leading eigenvalues (bars) and standard deviation over 5 repeated trainings (black lines). Mid and bottom rows: training points projected on the $\phi-\psi$ plane, colored according to the related eigenfunctions Deep-TICA 1 and 2. Remarkably, already after a couple of transitions the Deep-TICA procedure is able to extract a very good approximation of the eigenfunctions.}
\label{fig:si-ala2-training-configs}
\end{figure}
\vfill
\clearpage
\subsection*{Multithermal simulation - 2D free energy}
~

\vfill
\begin{figure}[h!]
\if\mycmd0
  \captionsetup{font={normalsize},textfont={normalfont}}
\fi
\includegraphics[width=0.7\linewidth]{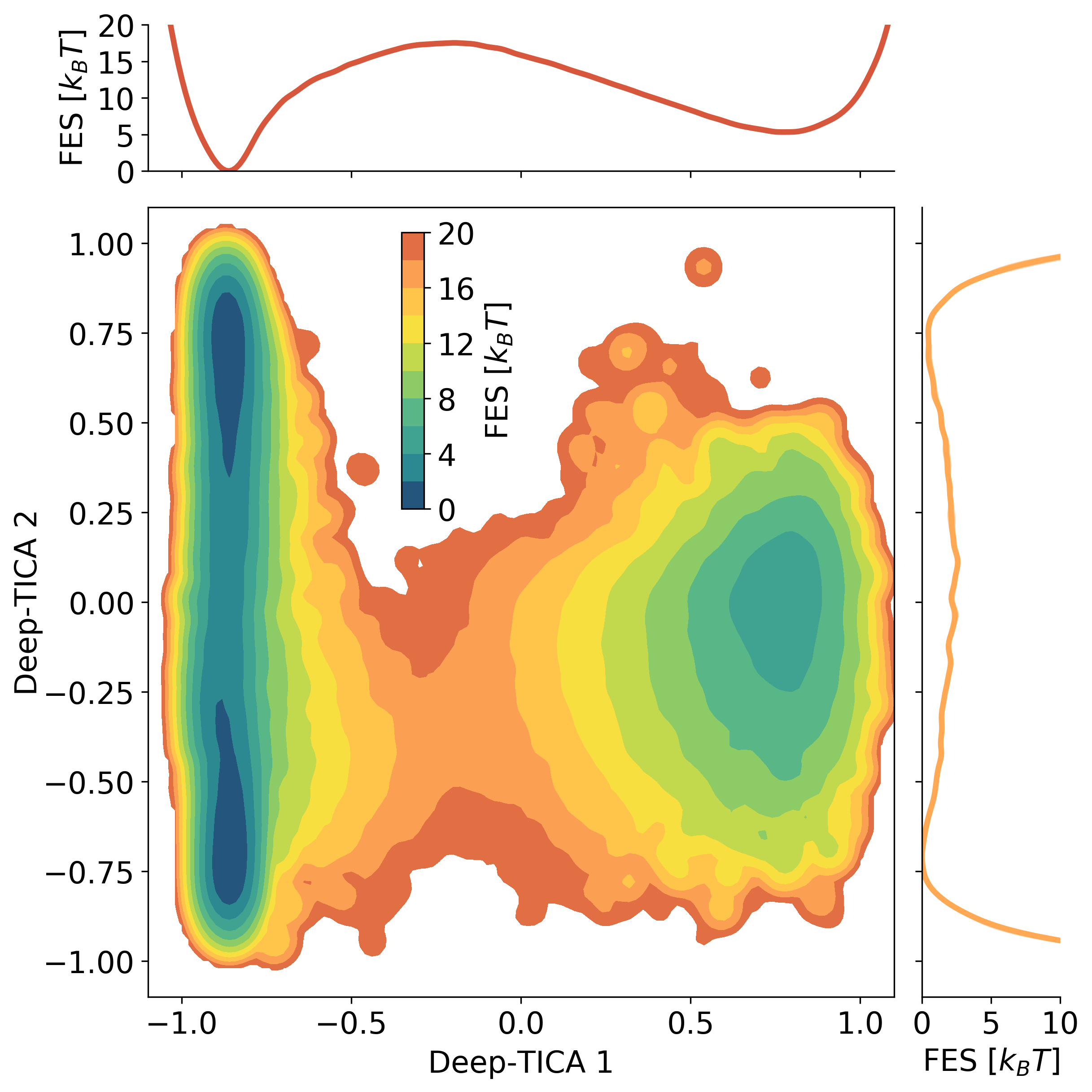}
\centering
\caption{\textbf{Free energy profile of alanine dipeptide as a function of Deep-TICA CVs} from the multithermal simulation (central panel) and 1D projections on the CVs (top and left panels). The leading Deep-TICA CV describes the conformational transition between $C_{7eq}$ and $C_{7ax}$, while the second CV highlights the presence of the two substates within $C_{7eq}$.}
\label{fig:si-ala2-fes}
\end{figure}
\vfill
\clearpage

\subsection*{ Multithermal simulation - Convergence and 1D free energy profiles}
~
\vfill
\begin{figure}[h!]
\if\mycmd0
  \captionsetup{font={normalsize},textfont={normalfont}}
\fi
\includegraphics[width=0.95\linewidth]{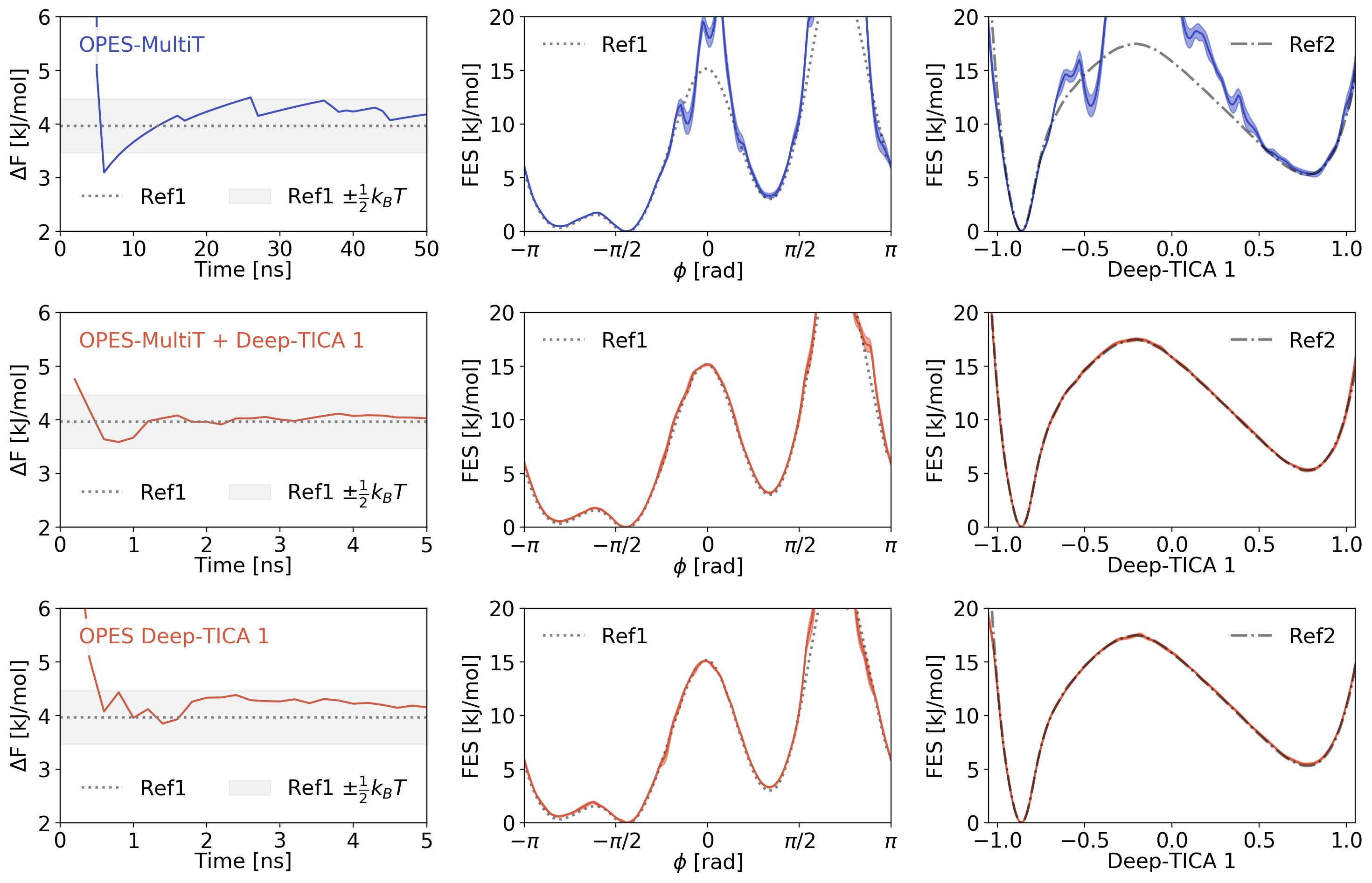}
\centering
\caption{\textbf{Convergence analysis} for the exploratory multithermal simulation (blue, first row), for the one biasing the Deep-TICA 1 CV in the multithermal ensemble (red, second row) and for an additional simulation in which only a bias potential along Deep-TICA 1 is applied with OPES (red, third row). In the first column we show the free energy difference $\Delta F$\protect\footnotemark between $C_{7eq}$ and $C_{7ax}$ as a function of time. In the second and third column we report the free energy profile along the $\phi$ dihedral angle and the Deep-TICA 1 CV, respectively. Blue and red shaded areas indicate the statistical uncertainties from a weighted block average technique~\cite{Invernizzi2020UnifiedSampling}. Dotted grey lines represents the reference values, obtained from two different simulations: Ref1 is obtained by performing a 100 ns OPES simulation biasing $\phi-\psi$, while Ref2 correspond to the OPES-MultiT+Deep-TICA 1 CV extended for 50ns. In both simulations biasing Deep-TICA 1 the FES is reconstructed with very high accuracy already after a few nanoseconds.
}
\label{fig:si-ala2-multi-convergence}
\end{figure}
\footnotetext{The free energy difference between the two states is defined as follows:
$\Delta F = \frac{1}{\beta} \log \frac{\int_A e^{-\beta F(s)} \mathrm{d} s }{\int_B e^{-\beta F(s)} \mathrm{d}s}$
where $s=\phi$ is the dihedral angle and $F(s)$ is the free energy profile. The two integrals are computed over the regions corresponding to $A=C_{7eq}$ and $B=C_{7ax}$, in this case $\phi<0$ and $\phi>0$.}

\begin{table}[h!]
\if\mycmd0
  \captionsetup{font={normalsize},textfont={normalfont}}
\fi
\centering
\label{tab:ala2-multi-rates}
\begin{tabular}{@{}ccc@{}}
\toprule
Simulation & Average transition rate [ps]    \\ \midrule
OPES-MultiT & 3125\\
 \midrule
OPES-MultiT + Deep-TICA 1 &  25\\
 \midrule
OPES Deep-TICA 1 & 45\\
\bottomrule
\end{tabular}
\vspace{1em}
\caption{\textbf{Average transition rates} between $C_{7eq}$ and $C_{7ax}$, obtained by dividing the simulation time by the number of transitions. A transition is recorded whenever the running average (on a 2 ps window) of the Deep-TICA 1 CV goes below 0.5 or above 0.5. Only the part of the simulations where the bias potential is quasi-static is used, to avoid counting transitions promoted by a rapidly changing external potential. It should be noted that each simulation samples a different target distribution, and this could affect the transition rate.}
\end{table}

\vfill

\clearpage
\subsection*{ $\psi$-biased simulation  - Convergence and 1D free energy profiles}
~

\vfill
\begin{figure}[h!]
\if\mycmd0
  \captionsetup{font={normalsize},textfont={normalfont}}
\fi
\includegraphics[width=0.95\linewidth]{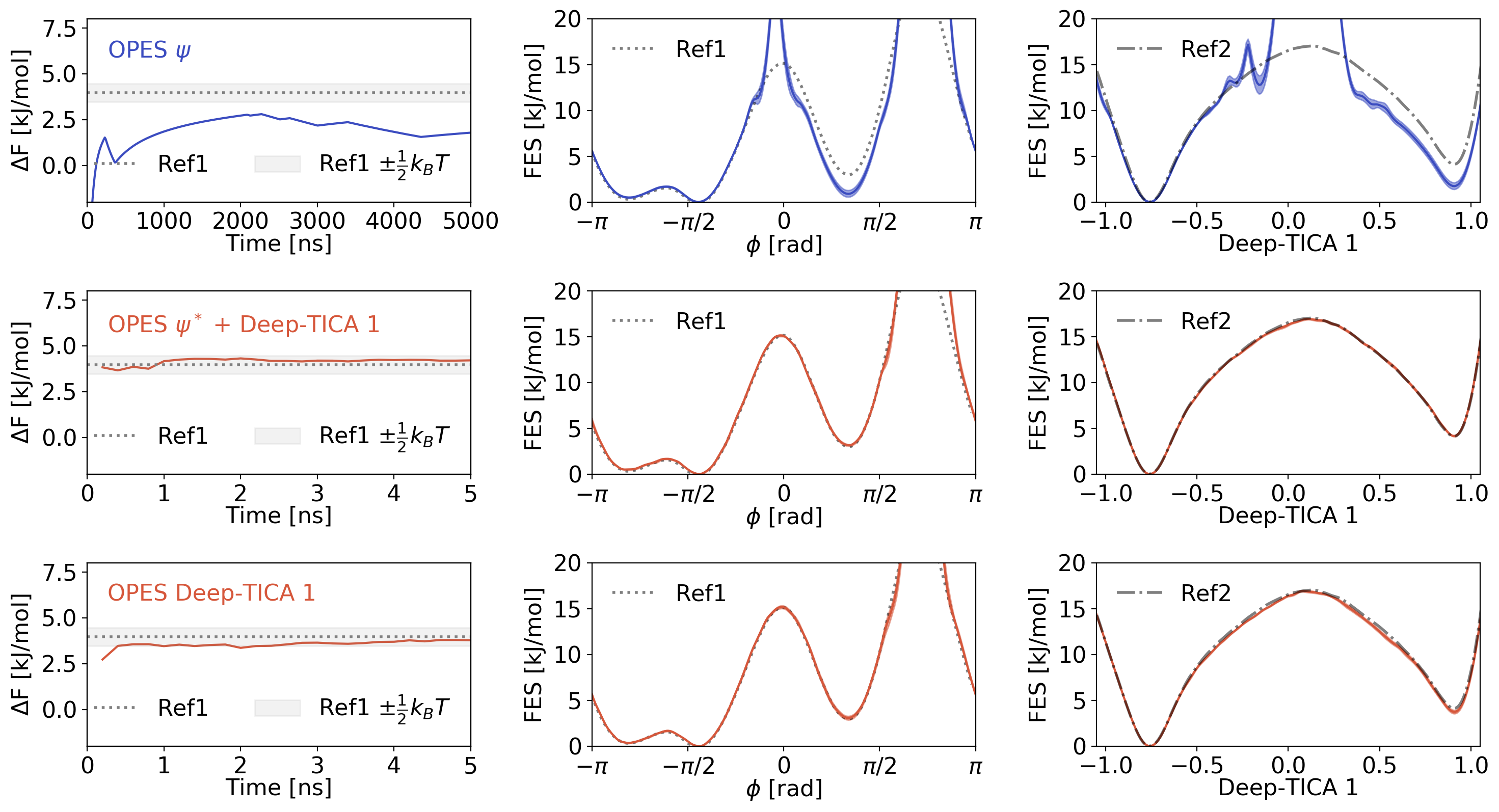}
\centering
\caption{\textbf{Convergence analysis} for the exploratory OPES simulation biasing $\psi$ (blue, first row), for the one biasing the Deep-TICA 1 CV together with the static bias $V^*(\psi)$ (red, second row) and for an additional simulation in which only a bias potential along Deep-TICA 1 is applied with OPES (red, third row). In the first column we show the free energy difference between $C_{7eq}$ and $C_{7ax}$ as a function of time. In the second and third column we report the free energy profile along the $\phi$ dihedral angle and the Deep-TICA 1 CV, respectively. Blue and red shaded areas indicate the statistical uncertainties from a weighted block average technique. Dotted grey lines represents the reference values, obtained from two different simulations: Ref1 is obtained by performing a 100 ns OPES simulation biasing $\phi-\psi$, while Ref2 correspond to the OPES $\phi$*+Deep-TICA 1 performed for 50ns. }
\label{fig:si-ala2-psi-convergence}
\end{figure}

\begin{table}[h]
\if\mycmd0
  \captionsetup{font={normalsize},textfont={normalfont}}
\fi
\centering
\label{tab:ala2-psi-rates}
\begin{tabular}{@{}ccc@{}}
\toprule
Simulation & Average transition rate [ps]    \\ \midrule
OPES $\phi$ & 455000\\
 \midrule
OPES $\phi$* + Deep-TICA 1&  20\\
 \midrule
OPES Deep-TICA 1 & 30\\
 \bottomrule
\end{tabular}
\vspace{1em}
\caption{\textbf{Average transition rates} between $C_{7eq}$ and $C_{7ax}$ for the enhanced sampling simulations. The values are obtained by dividing the simulation time by the number of transitions. A transition is recorded every time the running average (on a 2 ps window) of the Deep-TICA 1 CV goes below 0.5 or above 0.5. Only the part of the simulations where the bias potential is quasi-static is used, to avoid counting transitions promoted by a rapidly changing external potential. }
\end{table}

\vfill
\clearpage
\subsection*{Deep-TICA CVs comparison}
~

\vfill
\begin{figure}[h!]
\if\mycmd0
  \captionsetup{font={normalsize},textfont={normalfont}}
\fi
\includegraphics[width=0.9\linewidth]{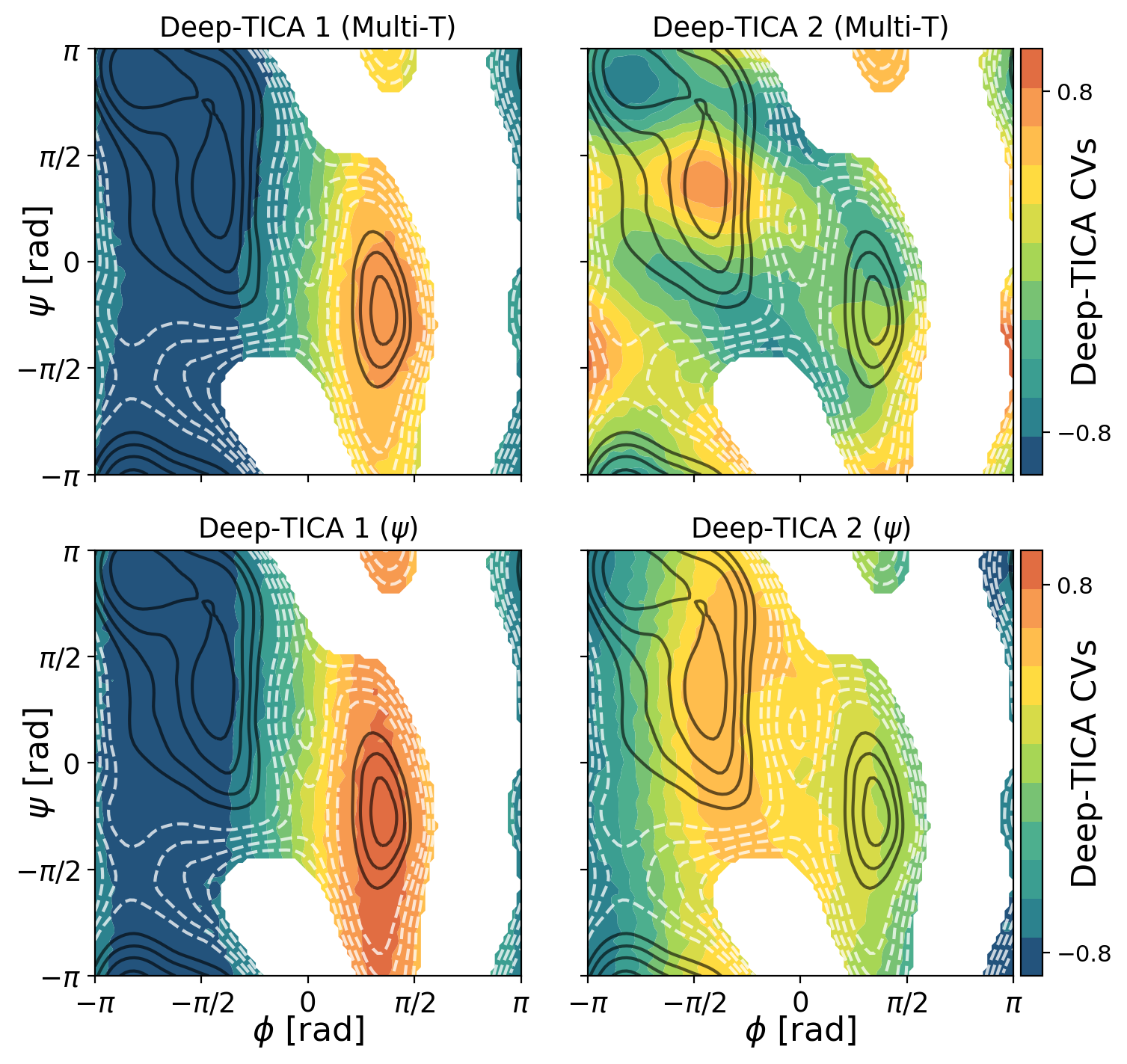}
\centering
\caption{\textbf{Deep-TICA CVs isolines in the Ramachandran plane}. The isolines have been computed from a 2D weighted histogram where the weights are the Deep-TICA CVs of the multithermal simulation (first column) and of the $\psi$-based OPES simulation (second column). In order to have a uniform sampling of this space, the configurations for the histogram are taken from a OPES simulation biasing $\phi$-$\psi$ with a flat target distribution. The two rows report the isolines of Deep-TICA 1 and 2, respectively. In addition we overlaid the isolines of the FES as a function of the Ramachandran angles, spaced every 2 $k_B T$ (white and black lines). In particular, solid black lines highlight the minima of the FES, while dashed white lines describe the higher energy regions. Only the regions where the FES $\leq$ 20 $k_BT$ are shown. It is noteworthy that in both the multicanonical and the $\psi$-biased dynamics the leading CV is associated to the transition between the $C_{7eq}$ and $C_{7ax}$ states, while the second one describes the transition between the two local minima separated by a much smaller free energy barrier within $C_{7eq}$. However, the isolines of the CVs extracted from the $\psi$-biased dynamics has no or little dependence on the $\psi$ angle, as the latter has made a fast degree of freedom in the original simulation.}
\label{fig:si-ala2-isolines}
\end{figure}
\vfill

\clearpage
\section*{B.CHIGNOLIN SIMULATIONS}
~
\subsection*{Replicas trajectories}
~
\vfill
\begin{figure}[h!]
\if\mycmd0
  \captionsetup{font={normalsize},textfont={normalfont}}
\fi
\includegraphics[width=\linewidth]{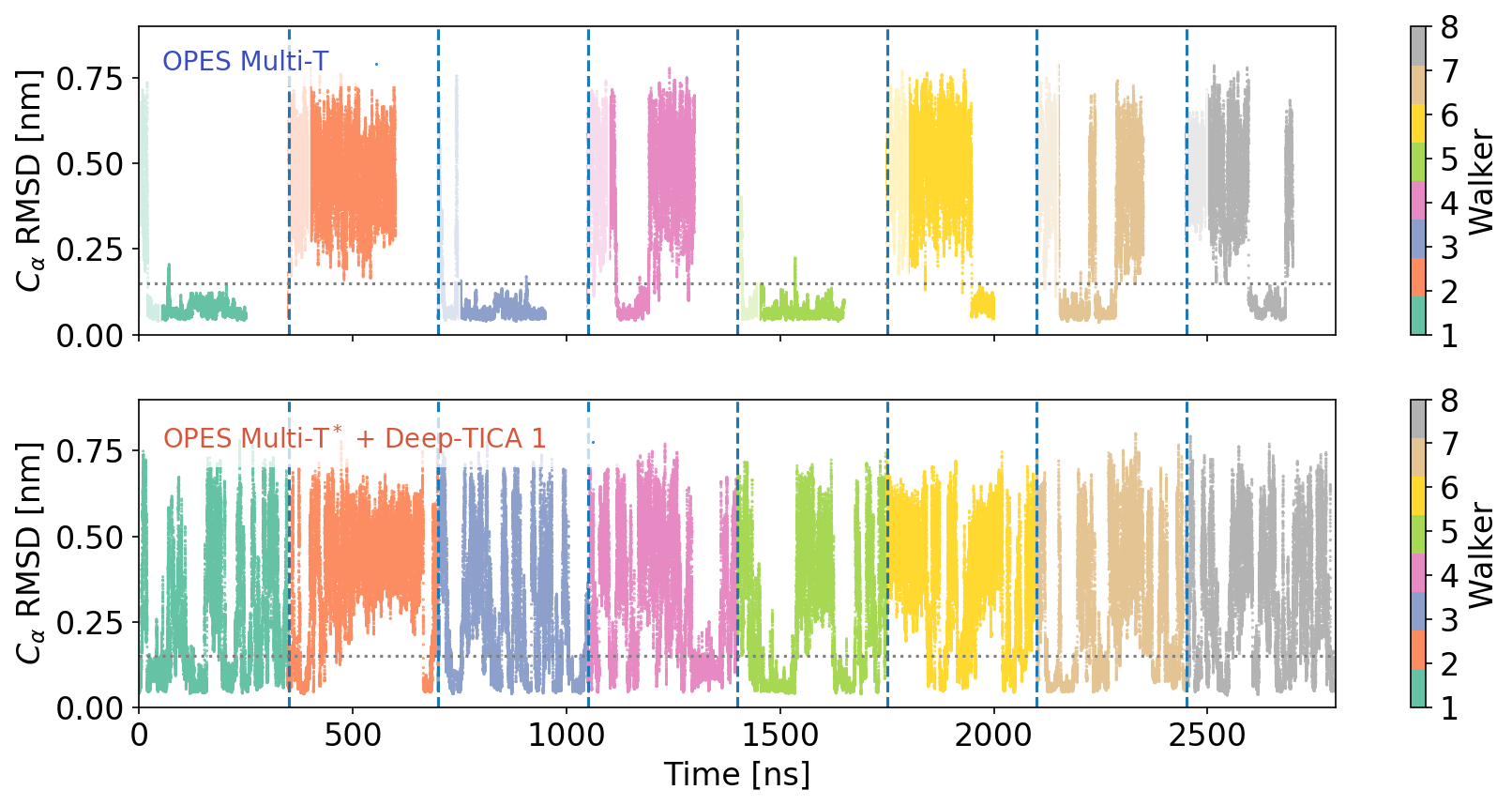}
\centering
\caption{\textbf{$C_\alpha$-RMSD time evolution} for the OPES Multi-T simulation (top) and for the OPES Multi-T* + Deep-TICA 1 run (bottom). Each color represent a replica of the system sharing the same bias potential. The simulation time is shifted by 350 ns times the replica id, and each replica is divided by the others by a vertical dashed line. In the case of the exploratory multithermal simulation the first 50 ns of each replica are made transparent to underline they are not used for the Deep-TICA training. A horizontal dotted line at RMSD=0.15 is added to identify folding-unfolding events.}
\label{fig:si-chignolin-traj}
\end{figure}
\vfill
\clearpage
\subsection*{Improving sampling of the transition region}
~
\vfill
\begin{figure}[h!]
\if\mycmd0
  \captionsetup{font={normalsize},textfont={normalfont}}
\fi
\includegraphics[width=0.9\linewidth]{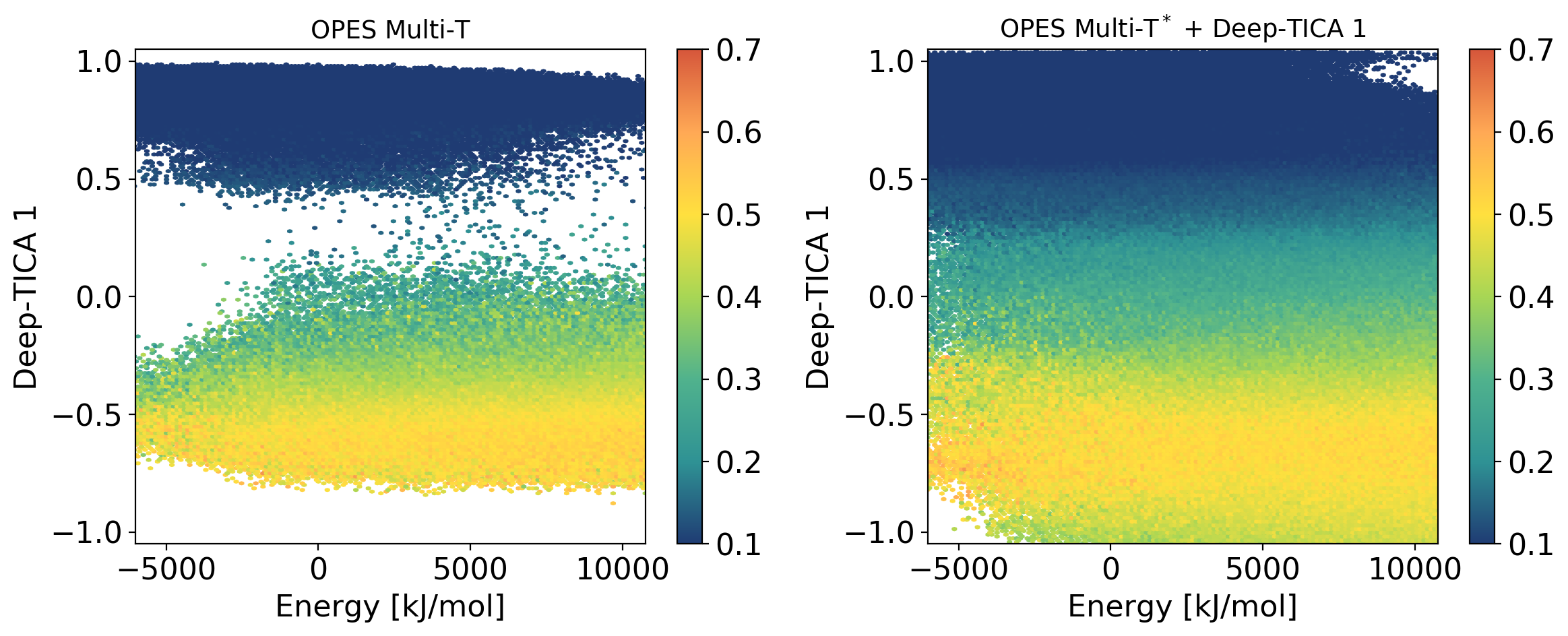}
\centering
\caption{\textbf{Scatter plot of the potential energy versus Deep-TICA 1} for the OPES Multi-T simulation (left) and for the OPES Multi-T* + Deep-TICA 1 run (right). The potential energy range corresponds to the temperature range [280K, 500K], and the energy is shifted such that the zero value corresponds to the average potential energy at 340K. The points are colored with the value of the $C_alpha$-RMSD. In the exploratory simulation, transitions between the folded and unfolded states are scarce especially at low values of the potential energy, which correspond to low temperature configurations. We observe that biasing also Deep-TICA 1 leads to increased sampling of the transition region in a uniform manner. This allows for more efficient recovery of the expectation values of observables at thermodynamic conditions further away from the simulated one.}
\label{fig:si-chignolin-ene-tica-scatter}
\end{figure}
\vfill
\subsection*{2D free energy surfaces - comparisons}
~
\vfill
\begin{figure}[h!]
\if\mycmd0
  \captionsetup{font={normalsize},textfont={normalfont}}
\fi
\includegraphics[width=\linewidth]{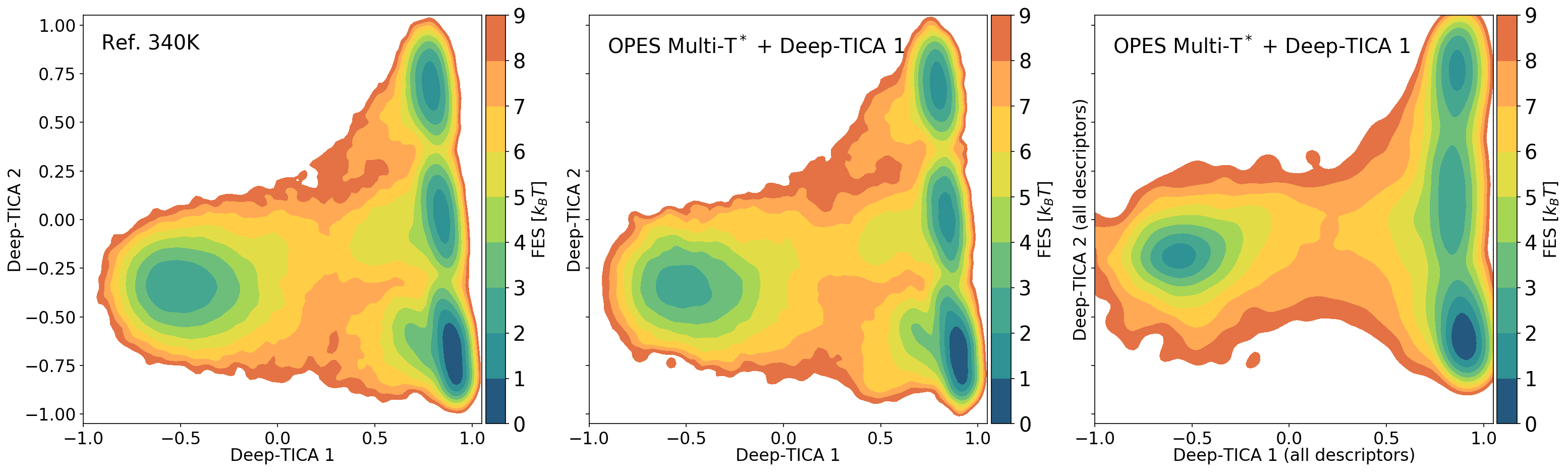}
\centering
\caption{\textbf{Comparison of the 2D free energies} as a function of the Deep-TICA CVs at T=340K. (left) Reference FES computed from the long unbiased run performed by Ref.~\cite{Lindorff-Larsen2011HowFold} (center) FES from OPES Multi-T* + Deep-TICA 1 simulation. (right) FES estimated from the same Deep-TICA simulation but projected on the Deep-TICA CVs trained with all the heavy atoms distances as input descriptors rather than on the limited subset. On one side, the agreement between the Deep-TICA simulation and the unbiased reference one is striking, providing evidence for the existence of these distinct folded metastable states. On the other side, the comparison with the CVs trained using all the distances confirms that the structure of the metastable states is not an artifact of using a reduced set of descriptors.}
\label{fig:si-chignolin-fes-comparison}
\end{figure}
\vfill
\clearpage
\subsection*{Free energy surface - temperature dependence}
~
\vfill
\begin{figure}[h!]
\if\mycmd0
  \captionsetup{font={normalsize},textfont={normalfont}}
\fi
\includegraphics[width=\linewidth]{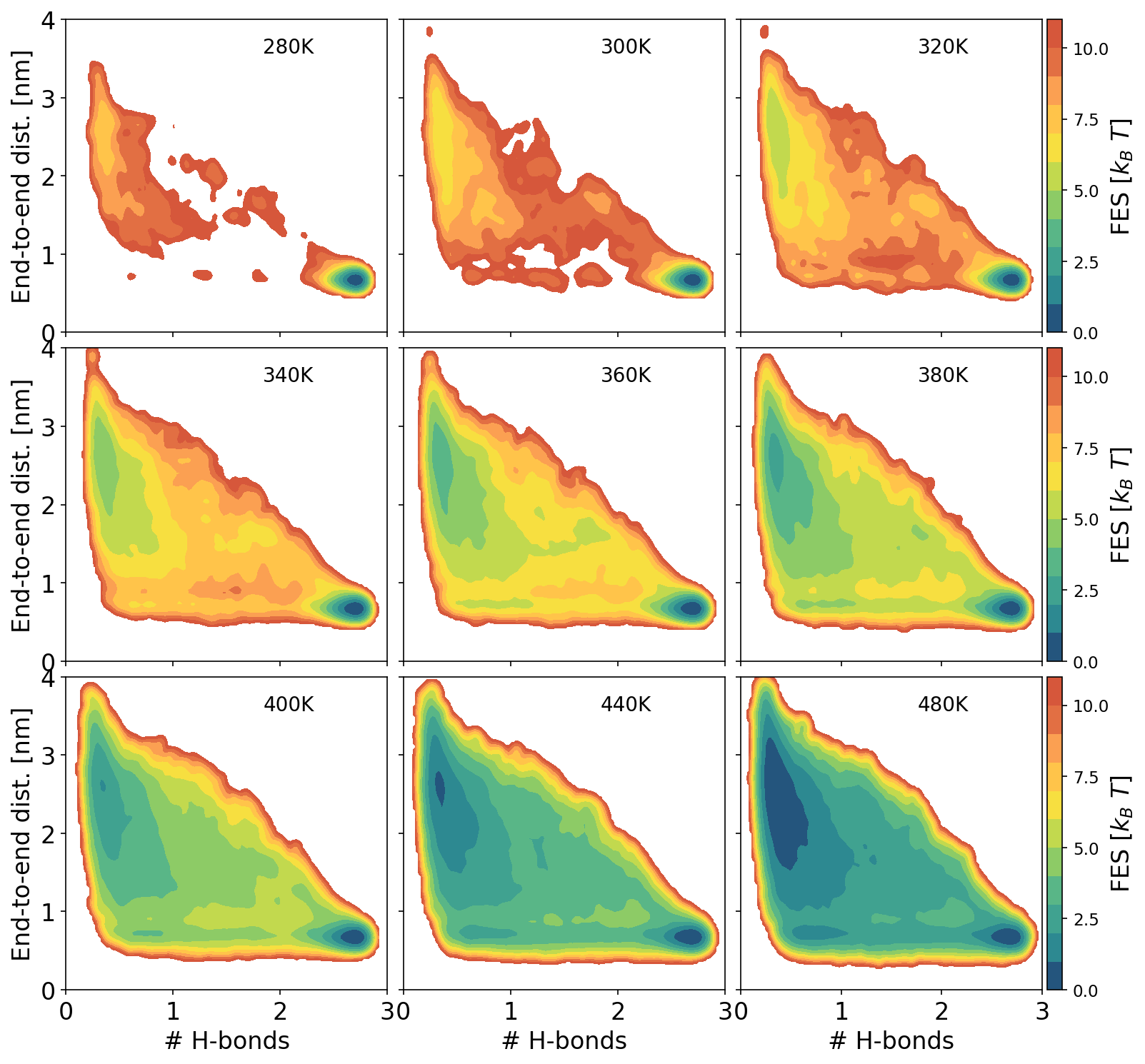}
\centering
\caption{\textbf{Temperature dependence of the FES} as a function of the number of H-bonds between backbone atoms \protect\footnotemark
and the end-to-end distance \protect\footnotemark. All the free energies are extracted from the single OPES Multi-T* + Deep-TICA 1 simulation. }
\label{fig:si-chignolin-fes2d-1-temp}
\end{figure}
\vfill
\footnotetext{The number of hydrogen bonds is calculated in PLUMED in a continuous way by summing the switching functions applied to the pairwise distances with parameters $R_0=4$\AA, N=6, M=8. The pair of atoms considered are: ASP3 N - THR8 O, GLY7 N - ASP3 O, TYR10 N - TYR1 O.}
\footnotetext{The end-to-end distance is computed by measuring the distance between the CA atoms of the two TYR terminal residues.}

\begin{figure}[h!]
\if\mycmd0
  \captionsetup{font={normalsize},textfont={normalfont}}
\fi
\includegraphics[width=\linewidth]{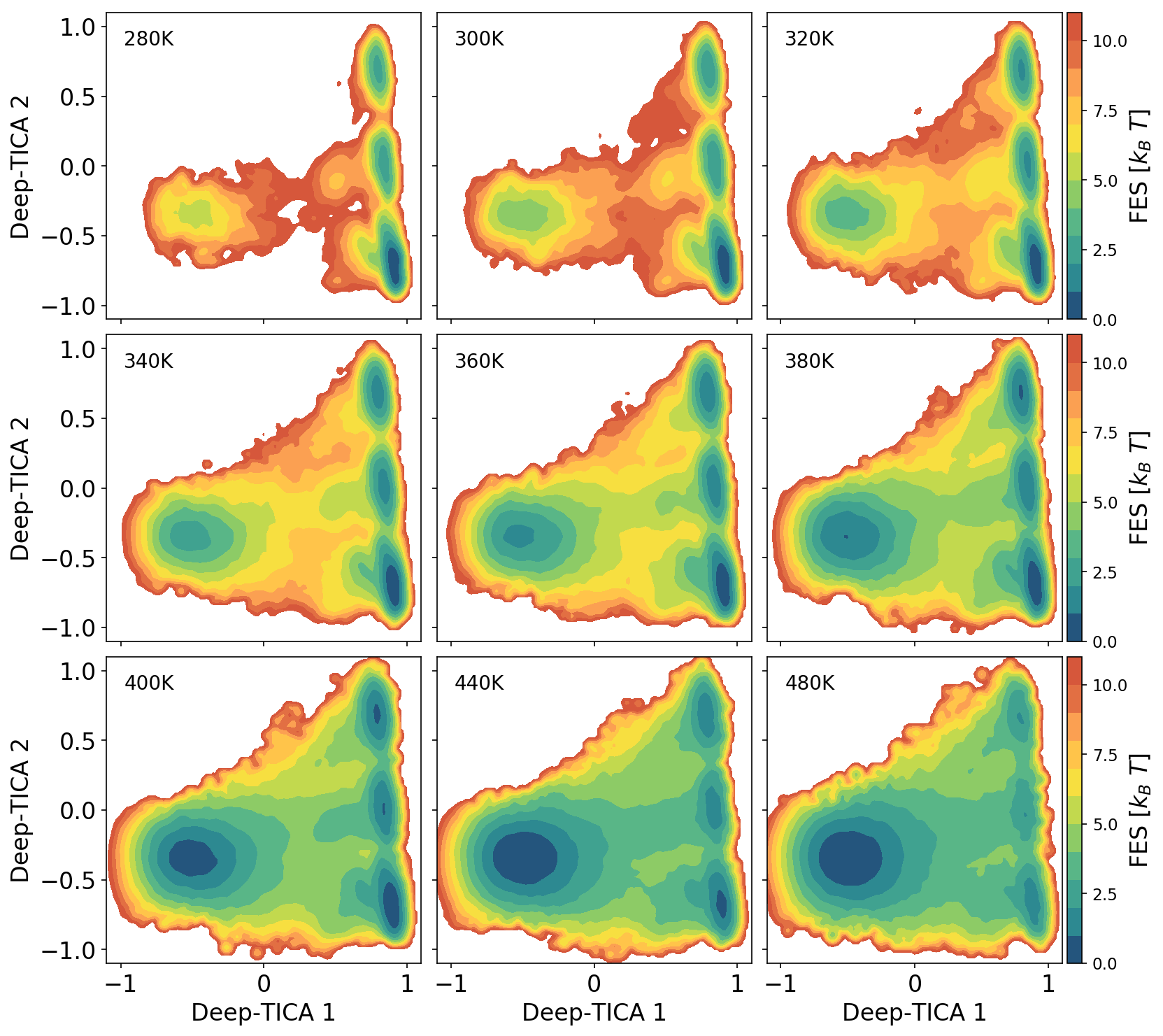}
\centering
\caption{\textbf{Temperature dependence of the FES as a function of the Deep-TICA CVs}, estimated from the OPES Multi-T* + Deep-TICA 1 simulation. }
\label{fig:si-chignolin-fes2d-2-temp}
\end{figure}

\begin{figure}[h!]
\if\mycmd0
  \captionsetup{font={normalsize},textfont={normalfont}}
\fi
\includegraphics[width=\linewidth]{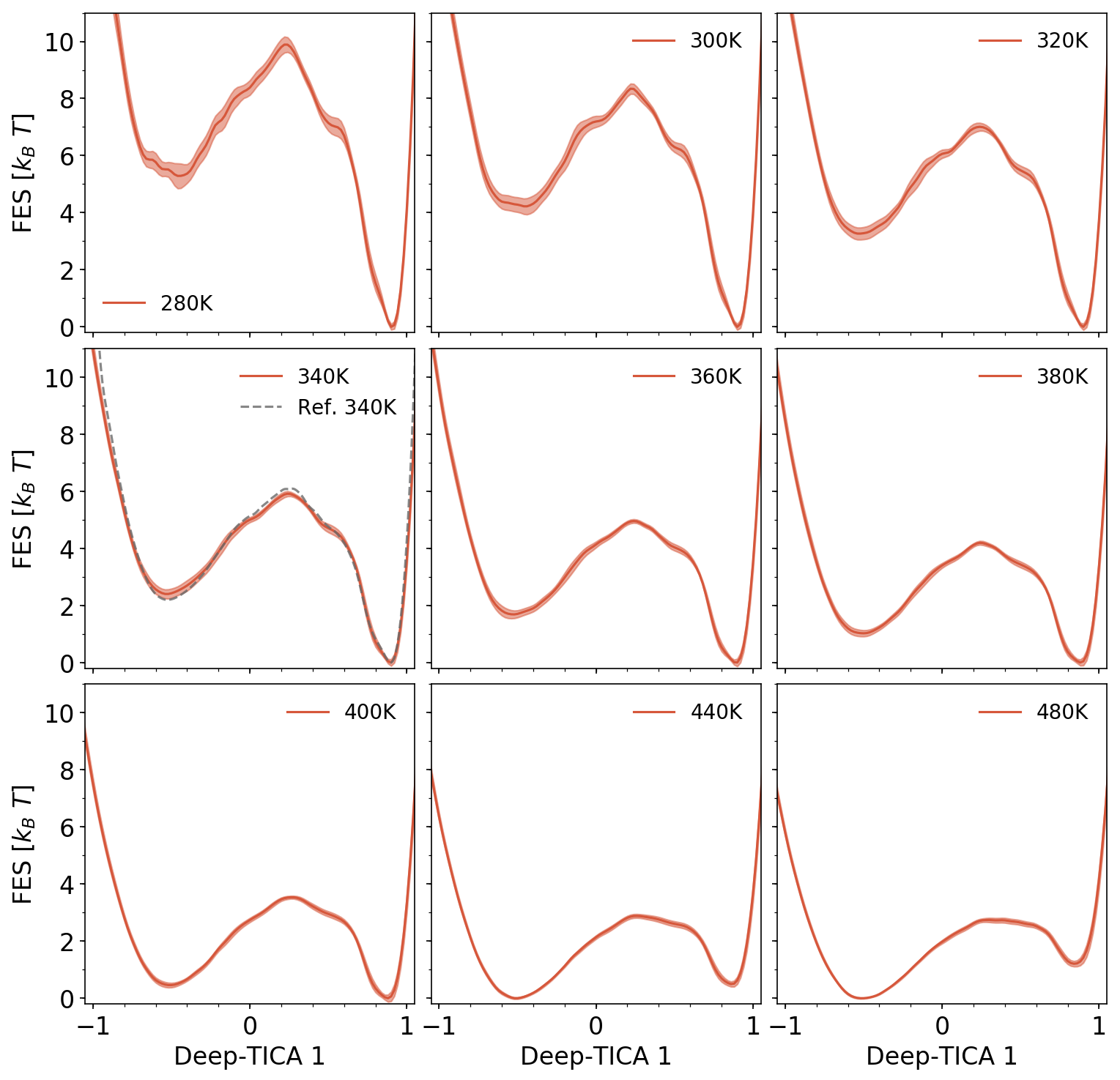}
\centering
\caption{\textbf{Temperature dependence of the FES as a function of Deep-TICA 1}, which describes the folding-unfolding transition, from the OPES Multi-T* + Deep-TICA 1 simulation. Shaded areas indicate the statistical uncertainty obtained with a weighted block average For the 340K temperature we report also the FES obtained from the reference unbiased simulation (grey dashed line).}
\label{fig:si-chignolin-fes1d-temp}
\end{figure}
\clearpage
\subsection*{Characterization of the folded states}
~

To extract the simulations corresponding to each folded basin we performed a weighted k-means clustering as in Fig.~4 in the manuscript, with weights $w=e^{\beta V(s)}$. Furthermore, a cutoff on the value of Deep-TICA 1>0.65 is used. Since none of the typical backbone descriptors (such as the $C_\alpha$-RMSD, the gyration radius or the end-to-end distance) is able to discriminate between the three folded states, we performed an hydrogen bonds analysis including also interactions with the side chains (Table~\ref{tab:hbonds}). 

\begin{table}[h]
\if\mycmd0
  \captionsetup{font={normalsize},textfont={normalfont}}
\fi
\centering
\begin{tabular}{@{}ccc@{}}
\toprule
States & Backbone & Sidechain     \\ \midrule
All & ASP3 N - THR8 O & THR6 N - ASP3 Os\\
& GLY7 N - ASP3 O & THR6 O1s - ASP3 Os\\
& TYR10 N - TYR1 O & TYR1 N - TYR10 Os\\
 \midrule
1 &  & THR8 Os - THR6 Os\\
 \midrule
2 &  & TYR10 Os - THR6 Os\\
 \midrule
3 &  & -\\
 \bottomrule
\end{tabular}
\vspace{1em}
\caption{Hydrogen bond analysis, performed using the hydrogen bonds command of
VMD~\cite{Humphrey1996VMD:Dynamics} with the same criterion used in Ref.~\cite{Maruyama2018AnalysisChignolin}, namely a cutoff distance = 3.3 \AA, an angle cutoff = 35° and a presence in at least 30\% of the configurations for each state.}
\label{tab:hbonds}
\end{table}

All these states have a common set of H-bonds, while differing in the presence or absence of specific bonds between the side chains. In particular, the most populated state (1) is characterized by the presence of an H-bond between the alcohol oxygens of the two threonine (THR) amminoacids. This implies that the interaction between the two sidechains have a stabilizing effect on the folded state, as it was observed in a structural analysis study for the wild-type chignolin protein~\cite{Maruyama2018AnalysisChignolin}. 
We can further characterize these states by looking at their correlations with the dihedral angles of the THR sidechains, from which we learn that state 2 can be further decomposed in two distinct states, characterized by different equilibrium values of the torsional angles. The emerging picture implies that there is not a single folded state, but rather an ensemble of folded structures.


\begin{figure}[h!]
    \centering
    \begin{minipage}{0.45\textwidth}
    \if\mycmd0
        \captionsetup{font={normalsize},textfont={normalfont}}
    \fi
        \centering
        \includegraphics[width=\textwidth]{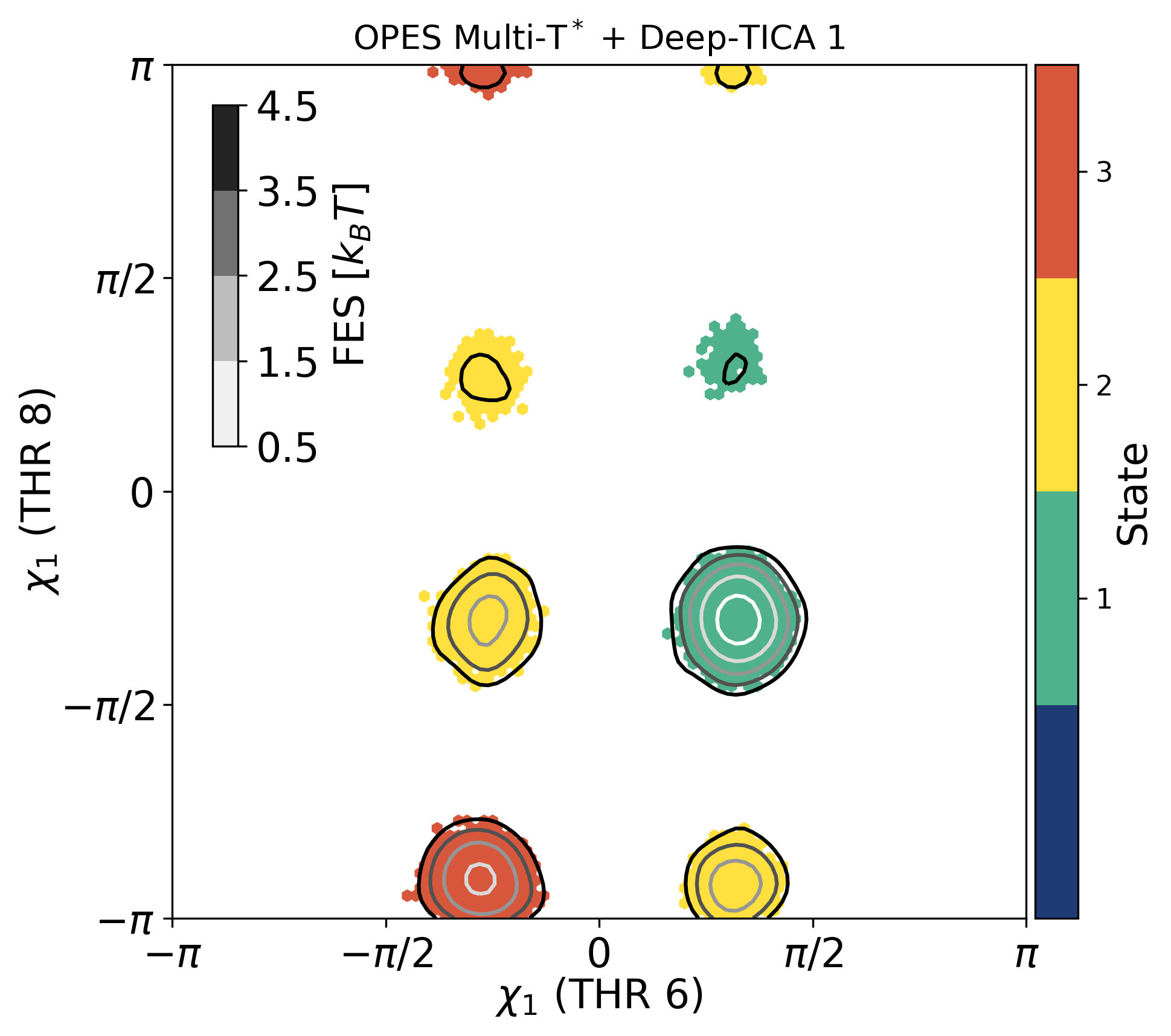} 
        \caption{\textbf{Scatter plot of the sidechain diedhral angles} $\chi_1$ (THR 6) and $\chi_1$ (THR 8) for the three folded states. The isolines of the FES at T=340K are also reported (solid lines, the color denotes the FES value).}
       \label{fig:si-chignolin-sidechain}
    \end{minipage}\hfill
    \begin{minipage}{0.45\textwidth}
    \if\mycmd0
        \captionsetup{font={normalsize},textfont={normalfont}}
    \fi
        \centering
        \includegraphics[width=\textwidth]{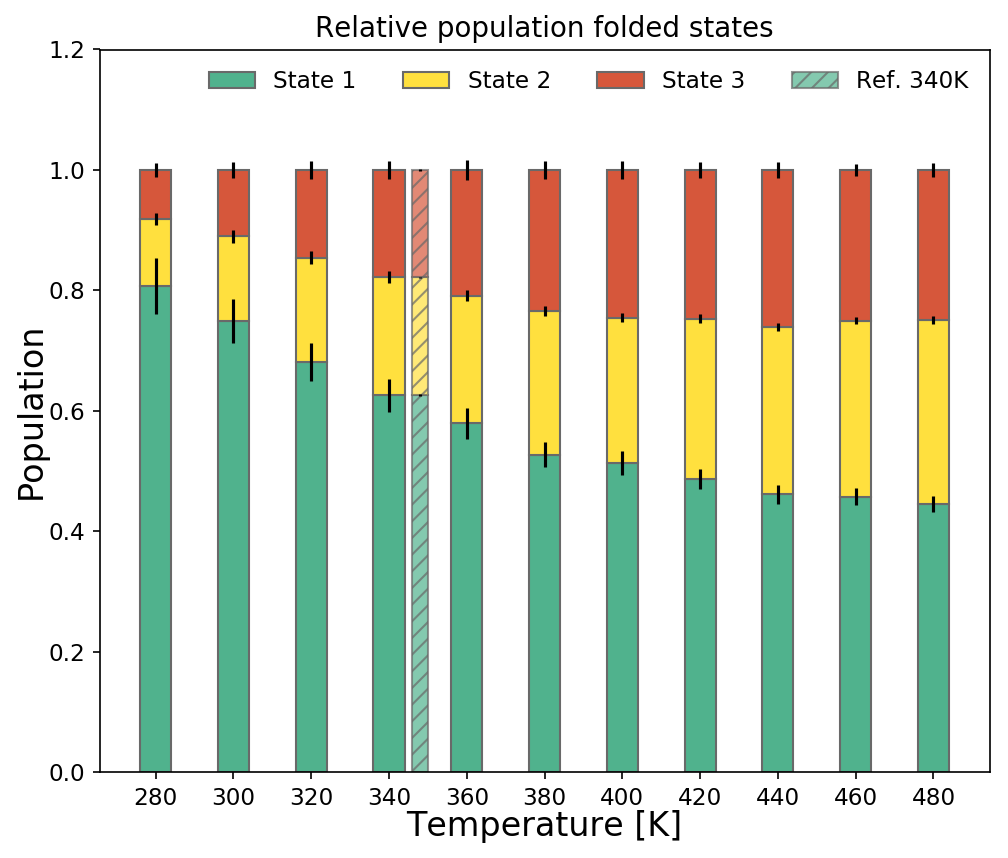} 
        \caption{\textbf{Relative population of the folded states as a function of temperature}, estimated by integrating the probability ditribution $P(s)=e^{-\beta F(s)}$ in each basin where s=Deep-TICA 2 and F(s) is obtained as in Fig.~5 by integration of the 2D FES with the condition Deep-TICA 1>0.65. } 
        \label{fig:si-chignolin-relative-population}
    \end{minipage}
\end{figure}

\clearpage
\section*{C. SILICON SIMULATIONS}
\if\mycmd1
~
\fi
\subsection*{Structure factor-based descriptors}
~
\begin{figure}[h!]
\if\mycmd0
  \captionsetup{font={normalsize},textfont={normalfont}}
\fi
\includegraphics[width=0.6\linewidth]{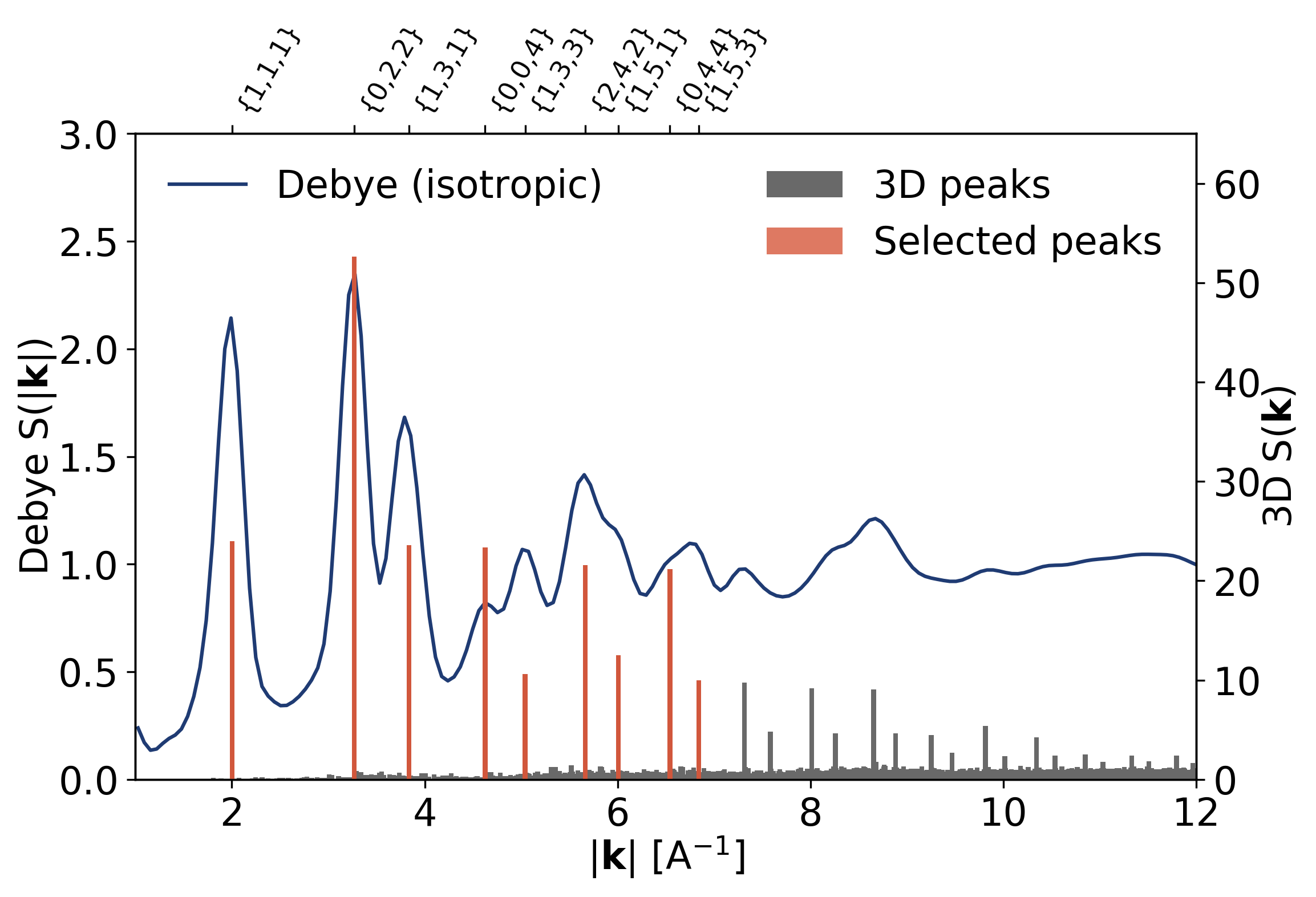}
\centering
\caption{\textbf{Structure factor peaks for cubic diamond crystal structure}. Bars correspond to the average value of three dimensional structure factor peaks as a function of $|\bm{k}|$ (right y-scale). We also report the isotropic structure factor calculated with the Debye formula (left y-scale) for comparison. At variance with Debye $S(k)$, the 3D $S(\bm{k})$ peaks are not rotationally invariant, but measure the presence of a crystal structure commensurate with the box size and aligned with the box axis. Selected peaks for the Deep-TICA training are highlighted in red, and the associated Miller indices are reported above the figure.}
\label{fig:si-silicon-peaks}
\end{figure}

\subsection*{Free energy profiles and free energy difference versus time}
\if\mycmd1
~
\fi
\begin{figure}[h!]
\if\mycmd0
  \captionsetup{font={normalsize},textfont={normalfont}}
\fi
\includegraphics[width=0.75\linewidth]{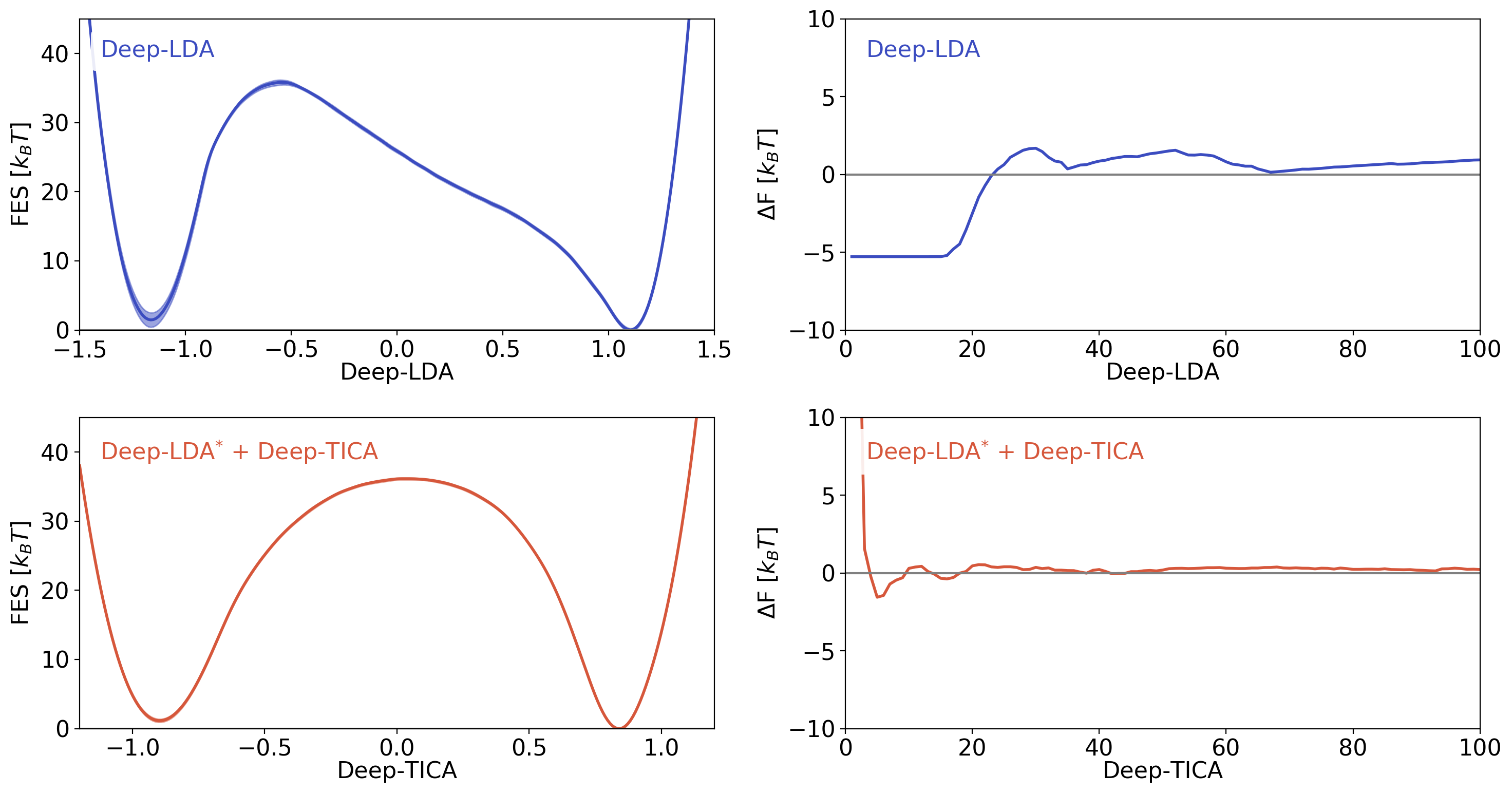}
\centering
\caption{\textbf{Free energy profiles} for the Deep-LDA simulation (top row) and the Deep-TICA simulation (bottom row). In the left column we report the FES as a function of Deep-LDA and Deep-TICA CV, respectively. Shaded areas correspond to the statistical uncertainties estimated with a weighted block average. In the right column we report the free energy difference versus time, estimated as in Fig.~\ref{fig:si-ala2-multi-convergence}}.
\label{fig:si-silicon-convergence}
\end{figure}

\clearpage



\end{document}